\documentclass{aa}
\usepackage{graphicx}
\usepackage{natbib}
\usepackage{amsmath}
\usepackage{amssymb}
\bibpunct{(}{)}{;}{a}{}{,}

\newcommand{\Ab}{\boldsymbol{A}}
\newcommand{\Cb}{\boldsymbol{{\mathrm C}}}
\newcommand{\Db}{\boldsymbol{D}}

\newcommand{\Ib}{\boldsymbol {I}}
\newcommand{\Jb}{\boldsymbol {J}}

\newcommand{\Nb}{\boldsymbol {N}}

\newcommand{\bb}{\boldsymbol{b}}

\newcommand{\nb}{\boldsymbol{n}}
\newcommand{\rb}{\boldsymbol{r}}

\newcommand{\xb}{\boldsymbol{x}}
\newcommand{\zb}{\boldsymbol{z}}
\newcommand{\oneb}{\boldsymbol{1}}
\newcommand{\zerob}{\boldsymbol{0}}

\newcommand{\gammab}{\boldsymbol{\gamma}}
\newcommand{\mub}{\boldsymbol{\mu}}
\newcommand{\sigmab}{\boldsymbol{\sigma}}

\newcommand{\nablab}{\boldsymbol{\nabla}}

\newcommand{\bh}{\widehat b}

\newcommand{\Ah}{\widehat A}

\newcommand{\Nh}{\widehat N}
\newcommand{\xh}{\widehat x}
\newcommand{\zh}{\widehat z}
\newcommand{\sigmah}{\widehat{\sigma}}
\newcommand{\gammah}{\widehat{\gamma}}

\newcommand{\Acb}{\boldsymbol{{\cal A}}}

\newcommand{\Deltab}{\boldsymbol{\Delta}}

\newcommand{\kmath}{\widehat {\cal K}}

\begin{document}

\title{Least-Squares methods with Poissonian noise: an \\ analysis and
a comparison with the Richardson-Lucy algorithm}

   \author{R. Vio\inst{1}
          \and
          J. Bardsley\inst{2}
          \and
          W. Wamsteker\inst{3}
          }

   \offprints{R. Vio}

   \institute{Chip Computers Consulting s.r.l., Viale Don L.~Sturzo 82,
              S.Liberale di Marcon, 30020 Venice, Italy\\
              ESA-VILSPA, Apartado 50727, 28080 Madrid, Spain\\
              \email{robertovio@tin.it}
         \and
                  Department of Mathematical Sciences, The University of Montana,
                        Missoula, MT 59812-0864, USA. \\
              \email{bardsleyj@mso.umt.edu}
        \and
             ESA-VILSPA, Apartado 50727, 28080 Madrid, Spain\\
             \email{willem.wamsteker@esa.int}
             }

\date{Received .............; accepted ................}

\abstract{It is well-known that the noise associated with the
collection of an astronomical image by a CCD camera is, in large
part, Poissonian. One would expect, therefore, that computational
approaches that incorporate this a priori information will be more
effective than those that do not. The Richardson-Lucy (RL)
algorithm, for example, can be viewed as a maximum-likelihood (ML)
method for image deblurring when the data noise is assumed to be
Poissonian. Least-squares (LS) approaches, on the other hand,
arises from the assumption that the noise is Gaussian with fixed
variance across pixels, which is rarely accurate. Given this, it
is surprising that in many cases results obtained using LS
techniques are relatively insensitive to whether the noise is
Poissonian or Gaussian. Furthermore, in the presence of Poisson
noise, results obtained using LS techniques are often comparable
with those obtained by the RL algorithm. We seek an explanation of
these phenomena via an examination of the regularization
properties of particular LS algorithms. In addition, a careful
analysis of the RL algorithm yields an explanation as to why it is
more effective than LS approaches for star-like objects, and why
it provides similar reconstructions for extended objects. We
finish with a convergence analysis of the RL algorithm. Numerical
results are presented throughout the paper. It is important to
stress that the subject treated in this paper is not academic. In
fact, in comparison with many ML algorithms, the LS algorithms are
much easier to use and to implement, often provide faster
convergence rates, and are much more flexible regarding the
incorporation of constraints on the solution. Consequently, if
little to no improvement is gained in the use of an ML approach
over an LS algorithm, the latter will often be the preferred
approach.

\keywords{Methods: data analysis -- Methods: statistical --
Techniques: Image processing} }

\titlerunning{Algorithms for Image Restoration}

\authorrunning{R. Vio, J. Bardsley, \& W. Wamsteker}

\maketitle

\section{Introduction}

The restoration of images is a common problem in Astronomy.
Astronomical images are blurred due to several factors: the
refractive effects of the atmosphere, the diffractive effects of
the finite aperture of the telescope, the statistical fluctuations
inherent in the collection of images by a CCD camera, and
instrumental defects. An illuminating example is represented by
the spherical aberration of the primary mirror of the  {\it Hubble
Space Telescope} \citep{whi91} that limited the quality of the
images before the detector system was refurbished.

The widespread interest in this subject has resulted in the
development of a large number of algorithms with different degrees
of sophistication \citep[for a review, see ][]{sta02}. For
example, recent wavelet-based approaches have been shown to
provide excellent results \citep[e.g., see ][]{nee04}.
Unfortunately, these algorithms are very expensive to implement,
prohibiting their use on large-scale image restoration problems
and on problems that require the restoration of a large number of
images. Consequently, for many restoration problems, less
sophisticated and computationally more efficient algorithms must
be used. In this respect, the algorithms based on a linear
Least-Squares (LS) methodology represent an interesting class. In
this paper we discuss two LS approaches: direct and iterative.
Direct methods, which we discuss in Sec. \ref{sec:tikhonov}, are
the most computationally efficient, while iterative techniques,
which we discuss in Sec. \ref{sec:itappr}, allow for the
straightforward incorporation of constraints.

In spite of the beneficial characteristics of the LS-based
algorithms, astronomers typically use techniques based on a
non-linear approach. Such algorithms are usually more difficult
to implement, are less flexible and often have slow convergence
rates. In particular, the original Richardson-Lucy (RL) algorithm
\citep{ric72, luc74, she82} and later modifications have seen a
great deal of attention. RL can be viewed as an
Expectation-Maximization (EM) algorithm associated with a Poisson
statistical noise model. Linear LS methods, on the other hand, can
be viewed as the maximum-likelihood (ML) approach when the noise
contaminating the image(s) of interest is additive Gaussian with
constant variance across pixels. For CCD camera noise, the
statistical assumption inherent in the RL algorithm is much more
accurate than that of the LS approach (see Sect.
\ref{sec:stat}). Nonetheless, it is often the case that these two
methods provide results of similar quality (see Sec.
\ref{sec:rl}). From a particular point of view, this fact is
disappointing, since it means that in certain instances the RL
algorithm is not able to exploit the a priori statistical
information. This is particularly relevant when the incorporation
of the a priori information results in algorithms that are more
expensive and difficult to implement.

The aims of the present paper are as follows: I) to determine the
performance of the LS algorithms when the noise is predominantly
Poissonian; II) to determine when the LS and RL approaches will
give very similar qualitative results. We stress that such
questions are not only of academic interest. In fact, the authors
believe that due to certain distinct computational advantages, the
LS algorithms should be used whenever their use is warranted. On
the other hand, we caution that it is {\em not} our intention to
conclude that the LS approach is always the best choice. In fact,
as we will show, this is not the case.

In the next section, we present the statistical model for CCD
camera image formation as well as the approximate model that we
will use in the paper. After some preliminary considerations in
Sec. \ref{sec:stat}, in Sec. \ref{sec:ls} we will explore
the convergence properties of two LS approaches. We will also
discuss the performance of these algorithms on different objects.
Finally, in Sec. \ref{sec:rl} we will explore in detail the
convergence properties the RL algorithm. Throughout the paper we
will present numerical results. Finally, we present our conclusion
in Section \ref{sec:conclusions}.

\section{Statistical Considerations}
\label{sec:stat}

Astronomical image data is typically collected with a device known
as a CCD camera. The following statistical model
\citep[see ][]{sny93,sny95} applies to image data from a CCD
detector array:
\begin{equation} \label{eq:rv_data}
b(i,j)= n_{\rm obj}(i,j) + n_{\rm bck}(i,j)+n_{\rm rdn}(i,j).
\end{equation}
Here, $b(i,j)$ is the data acquired by a readout of the pixels of
the CCD detector array; $i,j = 1, 2, \dots, N$ (without loss of
generality, square images are considered); $n_{obj}(i,j)$ is the
number of object dependent photoelectrons; $n_{\rm bck}(i,j)$ is
the number of background electrons; and $n_{\rm rdn}(i,j)$ is the
readout noise. The random variables $n_{\rm obj}(i_1,j_1)$,
$n_{\rm bck}(i_1,j_1)$, and $n_{\rm rdn}(i_1,j_1)$ are assumed to
be independent of one another and of $n_{\rm obj}(i_2,j_2)$,
$n_{\rm bck}(i_2,j_2)$, and $n_{\rm rdn}(i_2,j_2)$ for $i_1,
j_1\neq i_2,j_2$.

For clearness of presentation, it is helpful to use matrix-vector
notation. We rewrite Eq.~(\ref{eq:rv_data}) as
\begin{equation}
\label{eq:rv_data2} \bb = \nb_{\rm obj}+ \nb_{\rm bck}+\nb_{\rm
rdn},
\end{equation}
where the vectors have been obtained by column stacking the
corresponding two-dimensional arrays. The random vector $\nb_{\rm
obj}$ has a Poisson distribution with Poisson parameter $\Ab \xb$,
where $\xb$ is the true image, or object, and $\Ab$ is a matrix
that corresponds to the point spread function (PSF). Depending
upon the assumed boundary conditions, its structure is typically
block-circulant or block-Toeplitz \citep[e.g., see
][]{vio03,vog02}; $\nb_{bck}$ is a Poisson random vector with a
fixed positive Poisson parameter $\beta$; and $\nb_{\rm rdn}$ is a
Gaussian random vector with mean $0$ and fixed variance
$\sigma^2_{\rm rdn}$.

In the sequel, we will use the following notation to denote
model~(\ref{eq:rv_data2}):
\begin{equation} \label{eq:stat} \bb =
{\rm Poisson}[\Ab\xb] + {\rm Poisson}[\beta \cdot \oneb] +
\Nb_{(0,\sigma_{rdn}^2)},
\end{equation}
where ${\rm Poisson}[\mub]$ denotes a Poissonian random vector
with mean $\mub$, whereas $\Nb_{(\mub,\sigmab^2)}$ represents a
Gaussian random vector with mean $\mub$ and variance $\sigmab^2$
(for iid entries, $\mub = \mu$ and $\sigmab^2 = \sigma^2$). If
$\sigma_{\rm rdn}^2$ is large, we have ${\rm
Poisson}[\sigma^2_{\rm rdn}] \approx N_{(\sigma^2_{\rm
rdn},\sigma^2_{\rm rdn})}$ \citep[see][]{fel71}, and hence, using
the independence properties of the random variables in
Eq.~(\ref{eq:rv_data}) we obtain the following approximation of
Eq.~(\ref{eq:stat}):
\begin{equation} \label{eq:stat2}
\bb = {\rm Poisson}[\Ab\xb + \oneb \cdot (\beta +\sigma^2_{\rm
rdn})]-\sigma^2_{\rm rdn}.
\end{equation}

In order to simplify the analysis that follows, we assume the
following simplified model
\begin{equation} \label{eq:stat3}
\bb={\rm Poisson}[\Ab \xb].
\end{equation}
The analysis is easily extended to the model given by
Eq.~(\ref{eq:stat2}). Furthermore, in regions of high intensity
$\Ab \xb \gg \beta,\sigma^2$, in which case model~(\ref{eq:stat2})
is well-approximated by model~(\ref{eq:stat3}).

A further useful approximation is possible if the
elements of $\bb$ are large.
In fact, in this case model~(\ref{eq:stat3}) can be well
approximated with
\begin{equation} \label{eq:stat4}
\bb = \Ab \xb + \zb,
\end{equation}
where $\zb$ is a zero mean Gaussian random vector
\begin{equation} \label{eq:stat5}
\zb \approx  \gammab \odot \Nb_{(0,1)}.
\end{equation}
Here symbol ``$ ~\odot~ $'' denotes Hadamard (element-wise)
multiplication, and $\gammab = (\Ab \xb)^{1/2}$. In other words,
through \eqref{eq:stat4}, the nonlinear noise model
\eqref{eq:stat3} can be approximated with a linear, additive, {\it
nonstationary}, Gaussian noise model. Our own numerical
experiments suggest that $b_i > 40$ is sufficient. This condition
is true in many situations of practical interest (recall that
$b_i$ is the number of photons detected by the $i$th pixel in the
CCD camera).

\begin{figure}
        \resizebox{\hsize}{!}{\includegraphics{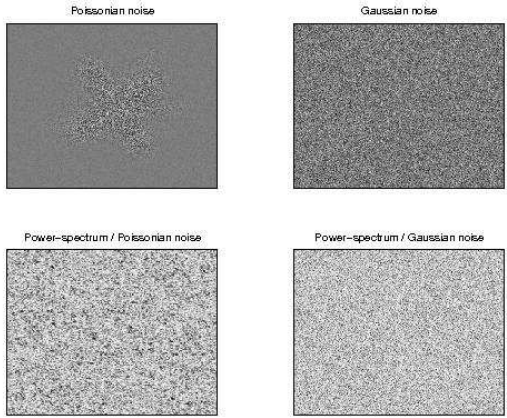}}
        \caption{Comparison of Poissonian vs. stationary Gaussian noise and corresponding
        power-spectra for the satellite image shown in Fig.~\ref{fig:gcv_sat40_1}.
        The variance of the two noises is the same.}
        \label{fig:noises}
\end{figure}

From equations~(\ref{eq:stat4}) and (\ref{eq:stat5}) we see that
$\zb$ has a flat spectrum, i.e., the expected modulus of its
Discrete Fourier Transform (DFT) is constant. However, here, at
difference with a gaussian white noise process, the various
Fourier frequencies are not independent from one another (e.g.,
see Fig.~\ref{fig:noises}). The reason is that the point-wise
multiplication in the spatial domain corresponds to convolution in
the Fourier domain, and vice versa. Thus, from \eqref{eq:stat5},
we have
\begin{equation} \label{eq:noise}
\zh(i,j) = [\gammah \otimes \Nh_{(0,1)}](i,j),
\end{equation}
where the symbol ``$~\widehat{}~$'' indicates DFT, ``$~\otimes~$''
denotes convolution, and $(i,j)$ represents a two-dimensional,
discrete frequency index. Since convolution is a linear operation,
and $\{ \Nh_{(0,1)}(i,j) \}$ are iid complex Gaussian random
variables with zero mean and unit variance, $\{ \zh(i,j) \}$ are
complex Gaussian random variables with zero mean and a constant
variance equal to $\sum_{i,j} |\gammah(i,j)|^2$.

\section{Performance of the Least-Squares Approach} \label{sec:ls}

The fact that the noise process $\zb$ has a flat spectrum,
provides some insight into the performance of the LS deblurring
algorithms in presence of Poissonian noise. In particular, since
the LS approach corresponds to the assumption that the
contaminating noise is an additive white noise process, the
possible worsening of the performance of LS algorithms has to be
expected due to their inability to take into account the
dependence between the Fourier frequencies that characterize the
spectrum of $\zb$. The question then arises, what happens if such
dependence is not taken into account? The following two arguments
suggest that, in many astronomical applications, the consequences
are not so important:
\begin{enumerate}
\item images of astronomical interest have spectra in which only
the lowest frequencies have intensities that are remarkably
different from zero. This is a consequence of the fact that the
PSFs are nearly band limited, i.e, they are very smooth functions.
Furthermore, if a function is in $\Cb^k$ (i.e., it has $k$
continuous derivatives) then its spectrum decreases at least as
fast as $1/\nu^{k+1}$. Consequently, this constitutes the
lower-limit with which the spectrum of the images can be expected
to decrease;

\item The discrete Picard's condition plus the Riemann-Lebesgue
lemma \citep{han97} show that the only Fourier frequencies useful
for the restoration of the image are (roughly) those where the
contribution of the object is larger than that of the noise.
\end{enumerate}
From such considerations, it is possible to conclude that in the
construction of the deblurred image only a few frequencies (i.e.,
the lowest ones) are primarily used, whereas most of the remaining
frequencies are of only marginal importance. For example, in the
case of a star-like source (i.e., a non-bandlimited function) and
Gaussian PSF with circular symmetry (a $\Cb^{\infty}$ function)
and dispersion (in pixels) $\sigma_p$, the observed spectrum is
again a Gaussian function with circular symmetry and dispersion
(in pixels) $\sigmah_p$ given by:
\begin{equation} \label{eq:sigmah}
\sigmah_p \approx N / (2 \pi \sigma_p).
\end{equation}
In several of the numerical experiments presented below, we have
used $N=256$ and $\sigma_p=12$. In this case, $\sigmah_p \approx
3.5$. With the levels used in the simulations, it happens that in
$\bh(i,j)$, the noise becomes dominant when (approximately) $i,j
\ge \gamma$, where $10\leq\gamma\leq 20$ corresponding to
different noise levels. Hence, the LS algorithms can be expected
to be almost insensitive to the nature of the noise.

Another point worth noting is that, although the set of
frequencies most corrupted by noise are determined both by the
noise level and by the spectrum of the PSF, it is the latter
factor that has the most influence. To see this we note that, for
the case of the star-like source and Gaussian PSF considered above, it
is possible to show that the frequency index $(i,j)$ where the
spectrum of the signal and that of $\zb$ have the same level is
given by $r \approx N \sqrt{\ln(A_s/\sigma_{\zb})} / (\pi \sigma_p
\sqrt{2})$, where $r=(i^2+j^2)^{1/2}$, $A_s$ is the amplitude of the
source, and $\sigma_{\zb}$ is the level of the
noise spectrum.

In the next two sections, we will check the reliability of the
above arguments in the context of two LS algorithms.

\subsection{The Tikhonov Approach}
\label{sec:tikhonov}

The linear systems that one encounters when solving image
restoration problems are often highly ill-conditioned. Because of
this, solutions can be extremely unstable. One of the most
standard approaches for handling this difficulty is known as
Tikhonov regularization. In the Tikhonov approach, one obtains a
stable estimate of the solution of the linear system of interest
by solving the penalized least squares problem
\begin{equation} \label{eq:tikhonov}
\xb_{\lambda} = {\rm argmin}\left(\, \Vert\, \Ab \xb - \bb
\,\Vert_2^2 + \lambda^2 \Vert\, \xb \,\Vert_2^2\, \right),
\end{equation}
or, equivalently, by solving the linear system
\begin{equation} \label{eq:tikhonov2}
(\Ab^T \Ab + \lambda^2 \Ib)\xb = \Ab^T \bb.
\end{equation}
Here $\lambda$ is a positive scalar known as the {\it
regularization parameter}. The direct solutions of
\eqref{eq:tikhonov2} can be obtained very efficiently using fast
implementations of the DFT. Moreover, there are various reliable and well
tested criteria that allow for the estimation of $\lambda$. A
standard technique is known as the generalized cross-validation
(GCV) method. With this approach, the optimal value of $\lambda$
is estimated via the minimization of the GCV function
\begin{equation}
\label{eq:gcv1}
{\rm GCV}(\lambda)= \frac{||\Ab\xb_\lambda - \bb||^{2}_{2}/ N^2}
 {[\,{\rm trace}(\Ib-\Acb(\lambda))/ N^2\,]^{2}}.
\end{equation}
Here $\Acb$ is the matrix that defines the estimator of $\Ab\bb$,
i.e., $\Acb\bb = \Ab\xb_{\lambda}$, and $N^2$ is the number of
pixels in the image. For Tikhonov regularization
\begin{equation}
\Acb(\lambda) =\Ab (\Ab^{T}\Ab+\lambda^{2}\Ib)^{-1}\Ab^{T}\bb.
\end{equation}

It is useful to express model~(\ref{eq:tikhonov}) and the GCV
function~(\ref{eq:gcv1}) in the Fourier domain:
\begin{equation} \label{eq:tikhonovf}
\xh_{\lambda}(i,j) = \frac{\Ah^*(i,j)}{| \Ah(i,j) |^2 + \lambda^2} \bh(i,j),
\end{equation}
and
\begin{align} \label{eq:gcvf}
  {\rm GCV}(\lambda)  = &
                 N^2
                 \sum_{i,j=0}^{N-1}
                 \left|\frac{\bh(i,j)}
                       {| \Ah(i,j) |^2+\lambda^{2}}
                 \right|^{2} \nonumber \\
                  & /
                  \left|
                         \sum_{i,j=0}^{N-1}
                         \frac{1}
                              {|\Ah(i,j)|^2+\lambda^2}
                  \right|^2.
\end{align}
Here, one can compute both $\xh_{\lambda}$ and the minimizer of
${\rm GCV}(\lambda)$ very efficiently.

\begin{figure}
        \resizebox{\hsize}{!}{\includegraphics{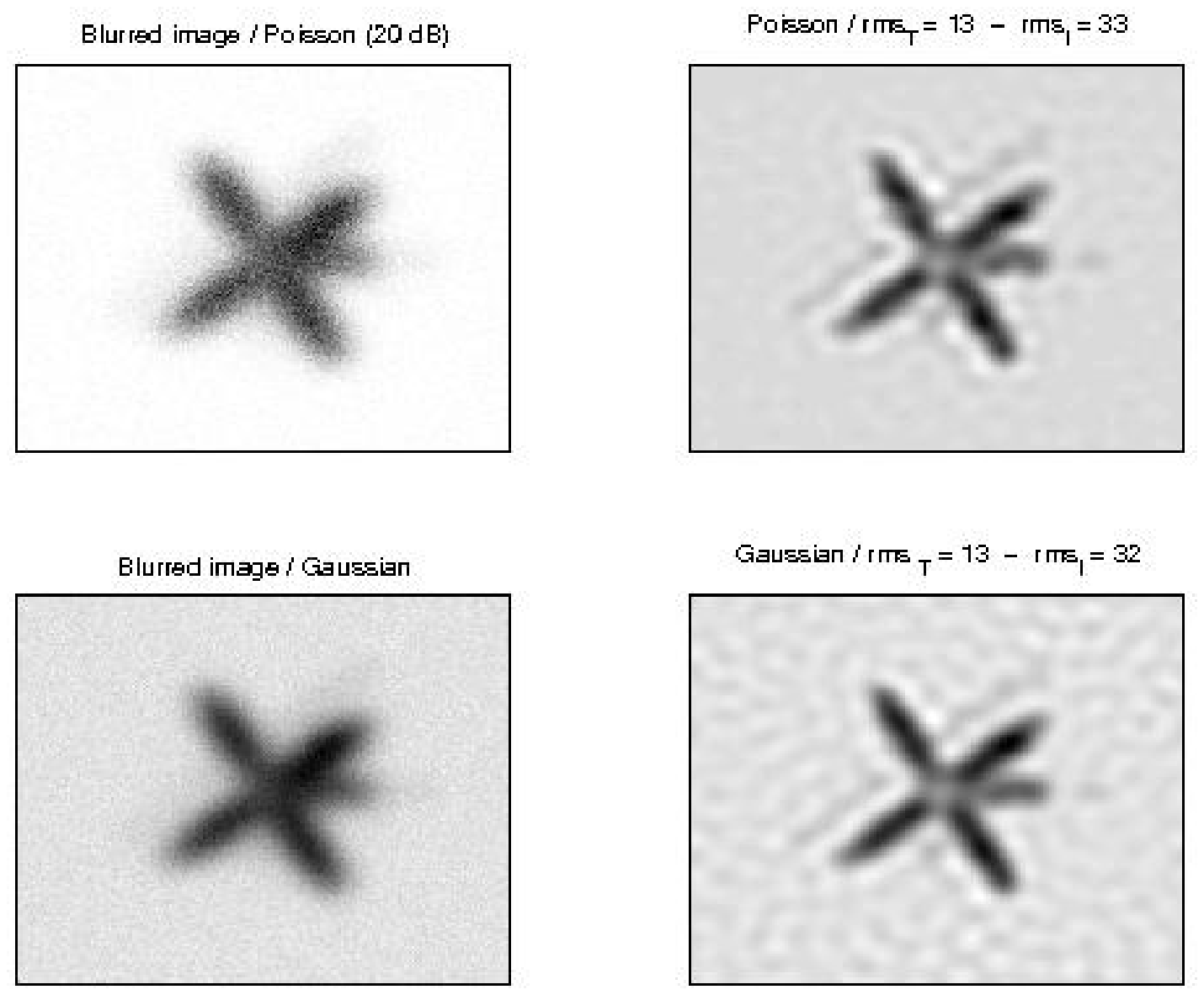}}
        \caption{Comparison of the results obtained by Tikhonov coupled with GCV in case of Poissonian
         and Gaussian noises (see text). The images have sizes $256 \times 256$ pixels, the PSF is Gaussian with
         circular symmetry and dispersion set to $12$ pixels, ${\rm S/N} = 20~{\rm dB}$.
         ${\rm rms}_T$ and ${\rm rms}_I$ are the rms of the true residuals calculated on the entire image and
         only on the pixels corresponding to the satellite, respectively.}
        \label{fig:gcv_sat20_1}
\end{figure}

\begin{figure}
        \resizebox{\hsize}{!}{\includegraphics{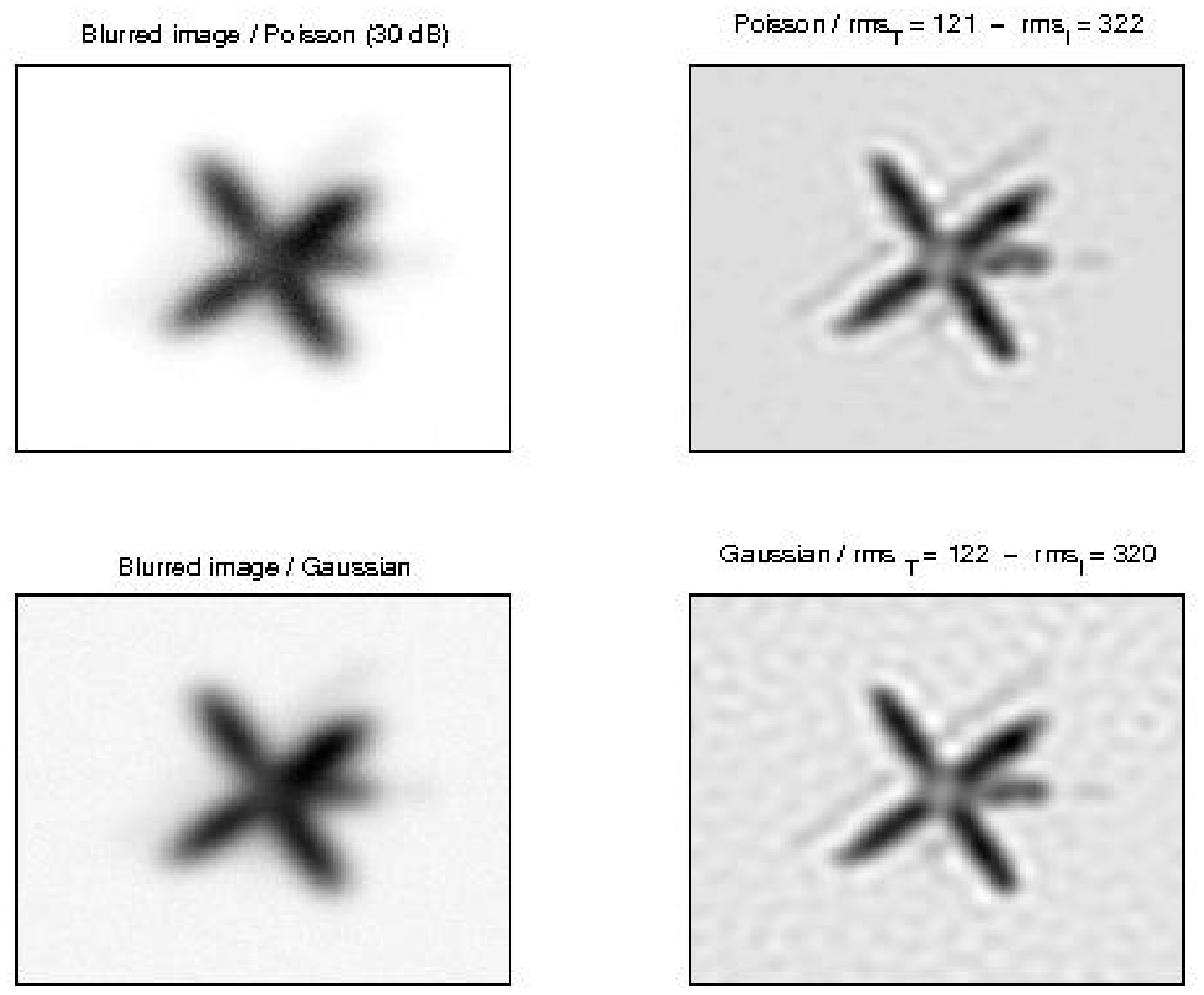}}
        \caption{Comparison of the results obtained by Tikhonov coupled with GCV in case of Poissonian
         and Gaussian noises (see text). The images have sizes $256 \times 256$ pixels, the PSF is Gaussian with
         circular symmetry and dispersion set to $12$ pixels, ${\rm S/N} = 30~{\rm dB}$.
         ${\rm rms}_T$ and ${\rm rms}_I$ are the rms of the true residuals calculated on the entire image and
         only on the pixels corresponding to the satellite, respectively.}
        \label{fig:gcv_sat30_1}
\end{figure}

\begin{figure}
        \resizebox{\hsize}{!}{\includegraphics{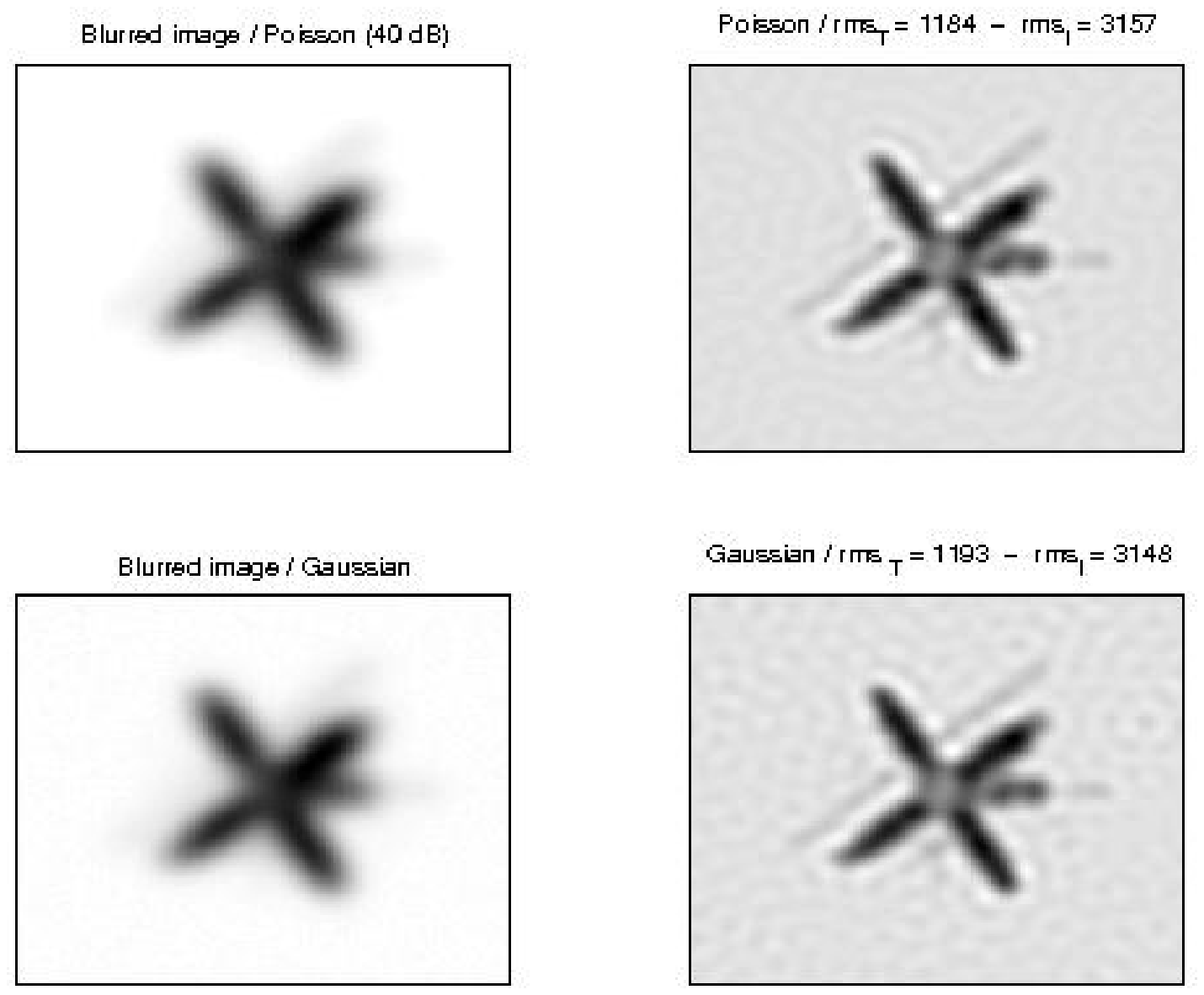}}
        \caption{Comparison of the results obtained by Tikhonov coupled with GCV in case of Poissonian
         and Gaussian noises (see text). The images have sizes $256 \times 256$ pixels, the PSF is Gaussian with
         circular symmetry and dispersion set to $12$ pixels, ${\rm S/N} = 40~{\rm dB}$.
         ${\rm rms}_T$ and ${\rm rms}_I$ are the rms of the true residuals calculated on the entire image and
         only on the pixels corresponding to the satellite, respectively.}
        \label{fig:gcv_sat40_1}
\end{figure}

\begin{figure}
        \resizebox{\hsize}{!}{\includegraphics{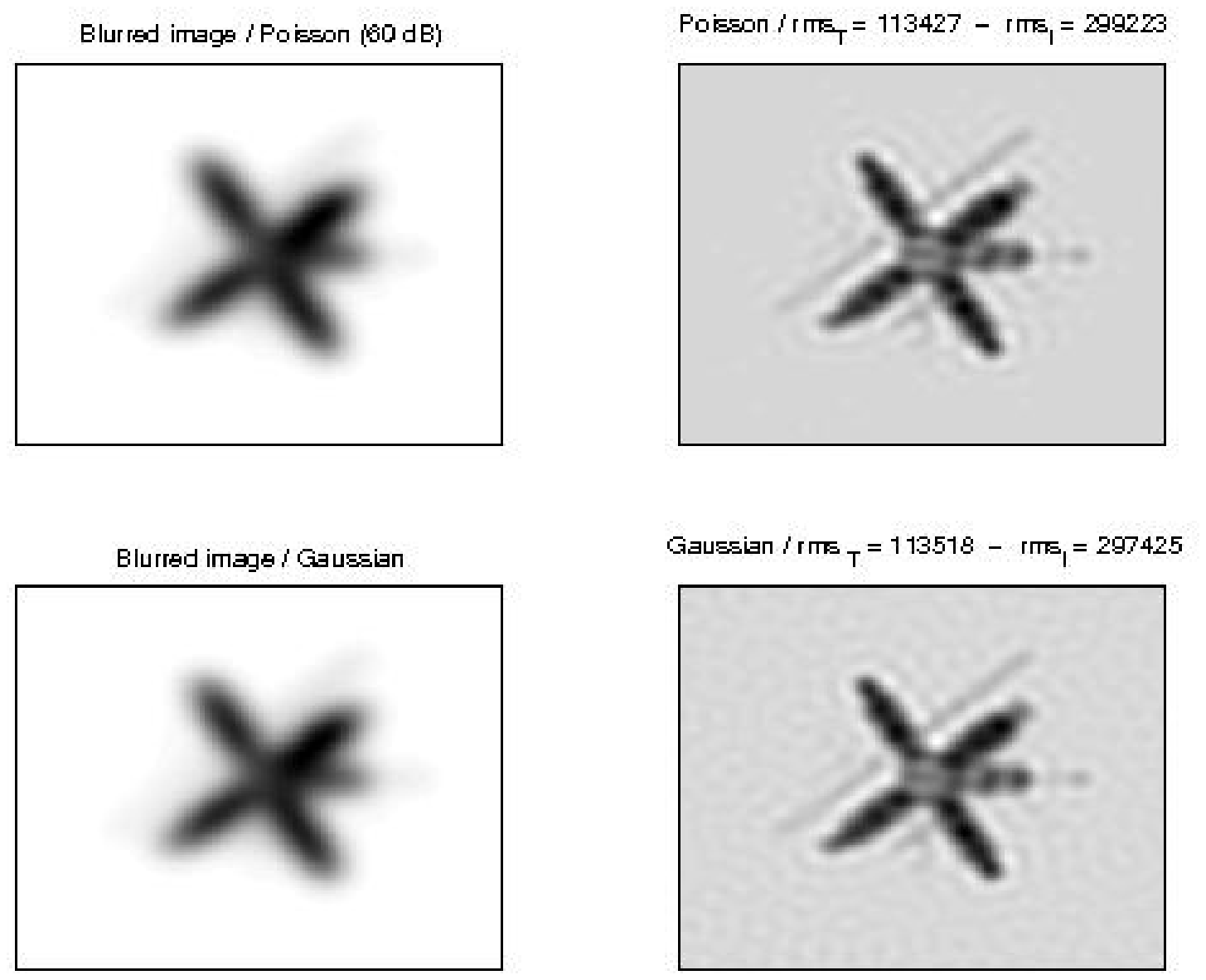}}
        \caption{Comparison of the results obtained by Tikhonov coupled with GCV in case of Poissonian
         and Gaussian noises (see text). The images have sizes $256 \times 256$ pixels, the PSF is Gaussian with
         circular symmetry and dispersion set to $12$ pixels, ${\rm S/N} = 60~{\rm dB}$.
         ${\rm rms}_T$ and ${\rm rms}_I$ are the rms of the true residuals calculated on the entire image and
         only on the pixels corresponding to the satellite, respectively.}
        \label{fig:gcv_sat60_1}
\end{figure}

Figures~\ref{fig:gcv_sat20_1} - \ref{fig:gcv_sat60_1} compare the
results obtainable with this method when noises are stationary
Gaussian and Poissonian, respectively. The image $b_p(i,j)$,
contaminated by Poissonian noise, has been obtained by simulating
a nonstationary Poissonian process with local mean given by the
values of the pixels in the blurred images, i.e. using
model~\eqref{eq:stat3}. Four peak signal to noise (${\rm S/N}$)
ratios have been considered \footnote{Here ${\rm S/N} =
20~\log(b_{{\rm max}} / b^{1/2}_{{\rm max}})~{\rm dB}$, where
$b_{{\rm max}}$ is the maximum value in the image $\Ab \xb$.}:
$20$, $30$, $40$ and $60~{\rm dB}$. They correspond to situations
of very low, intermediate, and very high noise levels. The PSF
used in the simulations is a two-dimensional Gaussian function
with circular symmetry. The image $b_g(i,j)$, contaminated by
Gaussian noise, has been obtained by the addition of a discrete
stationary white noise process to the blurred images. Both the
Gaussian and the Poissonian noises have been simulated through a
classic {\it inverse distribution method} \citep[e.g., see
][]{joh87} by transforming the same set of random uniform
deviates. They have exactly the same variance. Here, the subject
of interest is superimposed to a sky whose intensity, in the
blurred image, is set to $1\%$ of the maximum value of the image.
This means that, contrary to the Gaussian case where the noise
level is constant across the image, in the Poissonian case the
noise is mostly concentrated in the pixels with highest values. In
spite of this fact, these figures show that the results provided
by Tikhonov coupled with GCV are quite similar regardless of
whether the noise is of Gaussian of Poissonian type.

\begin{figure}
        \resizebox{\hsize}{!}{\includegraphics{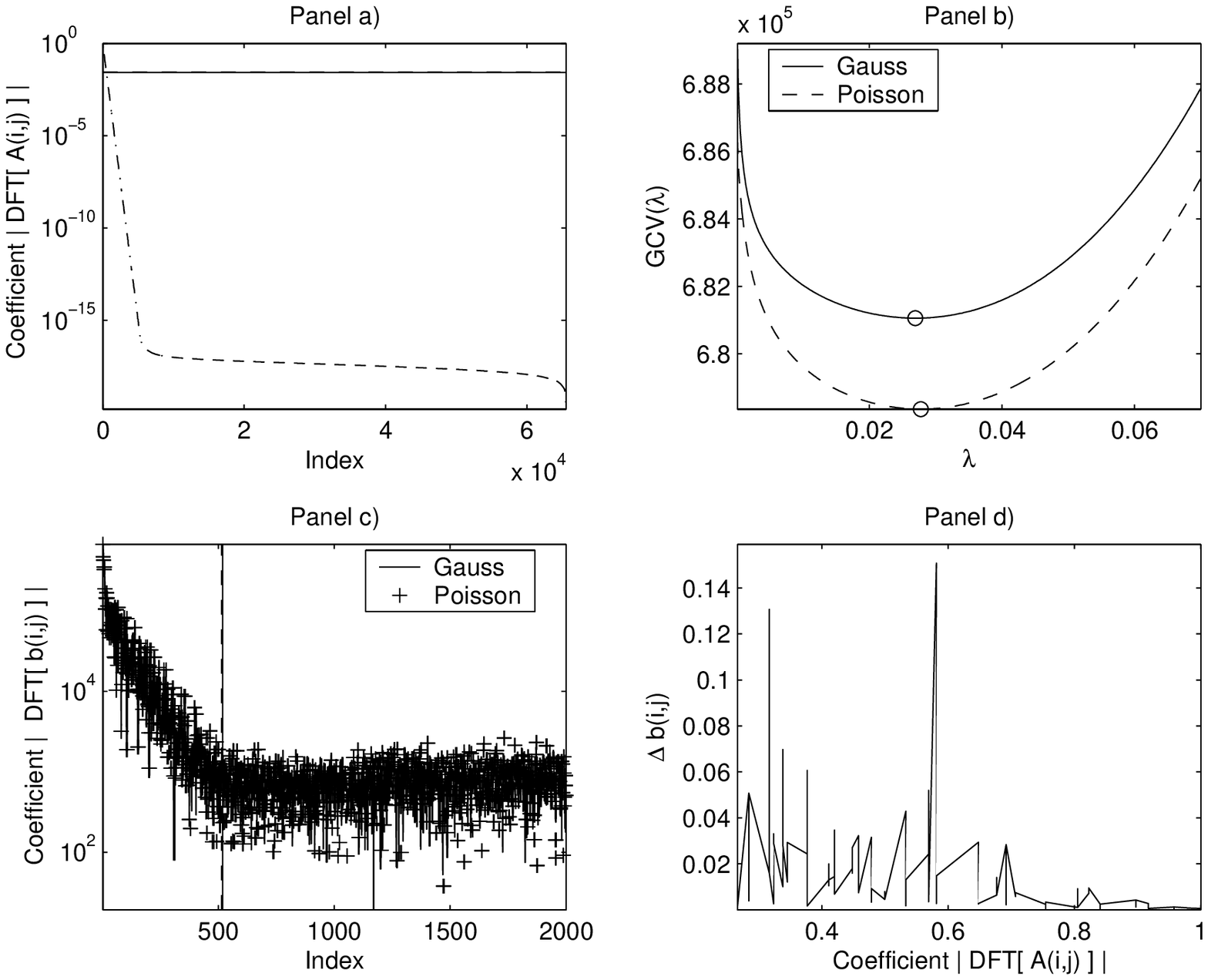}}
        \caption{Comparison of the results obtained by Tikhonov coupled with GCV in case of Poissonian
         and Gaussian noises, This figures correspond to the experiment shown in Fig.~\ref{fig:gcv_sat20_1}.
         Panel $a)$ the coefficients $| \bh_g(i,j) |$ and $| \bh_p(i,j) |$ in decreasing order.
        The two horizontal lines represent the values of $\lambda $ for the two noises; b) corresponding
        GCV functions; c) coefficients
        $|\bh_p(i,j)|$ and $|\bh_g(i,j)|$ corresponding to the first $2000$
        coefficients of $| \Ah(i,j) |$ shown in panel a). The vertical lines
        show the indices of $|\bh(i,j)|$ closest
        to $\lambda$. d) $\Delta b(i,j) = | \bh_g(i,j) - \bh_p(i,j) | / |\bh_g(i,j)|$ vs. the corresponding first $2000$
        coefficients of $| \Ah(i,j) |$ with the largest modulus. $ {\rm S/N} = 20~{\rm dB}$. }
        \label{fig:gcv_sat20_2}
\end{figure}

\begin{figure}
        \resizebox{\hsize}{!}{\includegraphics{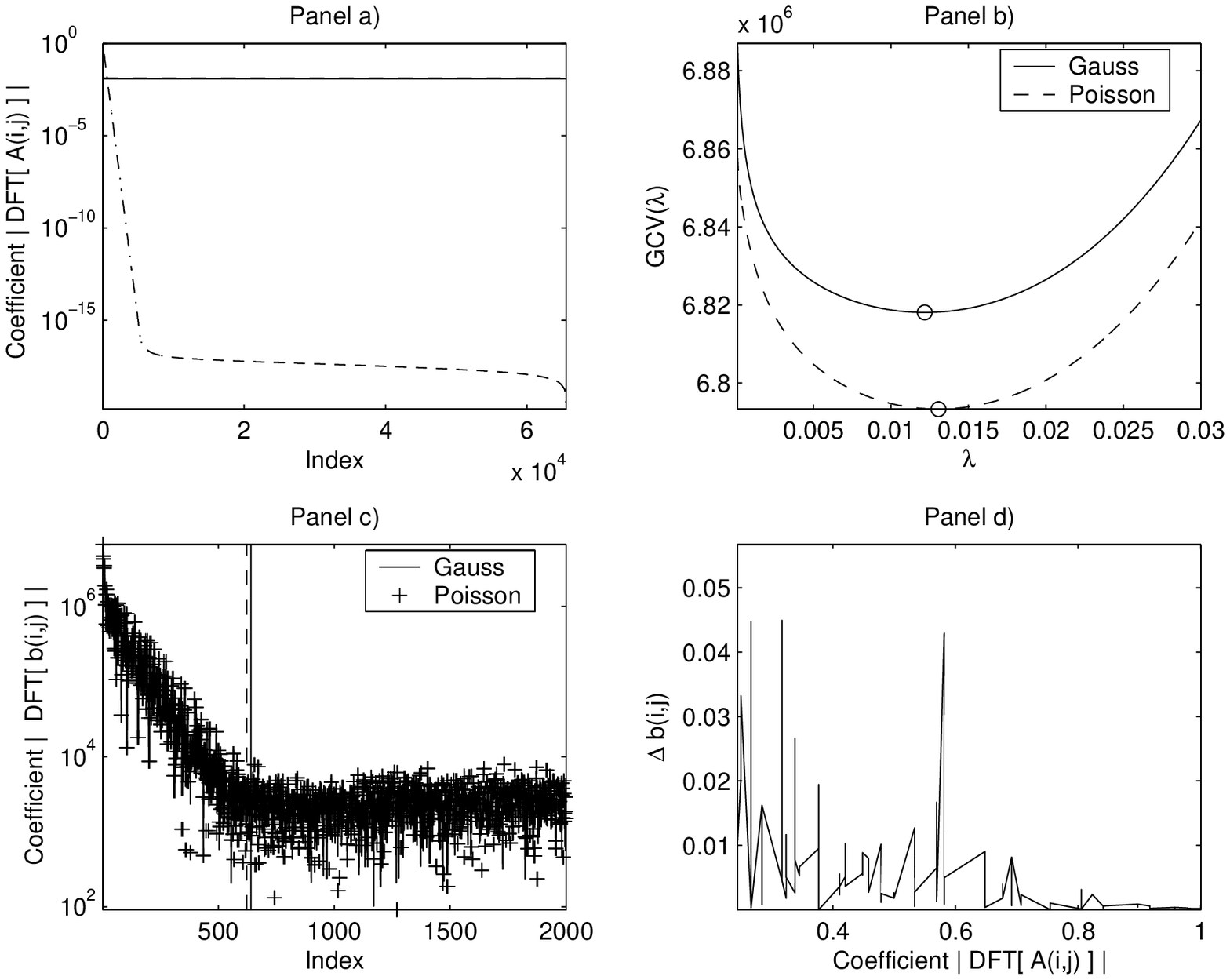}}
        \caption{Comparison of the results obtained by Tikhonov coupled with GCV in case of Poissonian
         and Gaussian noises, This figures correspond to the experiment shown in Fig.~\ref{fig:gcv_sat30_1}.
         Panel $a)$ the coefficients $| \bh_g(i,j) |$ and $| \bh_p(i,j) |$ in decreasing order.
        The two horizontal lines represent the values of $\lambda $ for the two noises; b) corresponding
        GCV functions; c) coefficients
        $|\bh_p(i,j)|$ and $|\bh_g(i,j)|$ corresponding to the first $2000$
        coefficients of $| \Ah(i,j) |$ shown in panel a). The vertical lines
        show the indices of $|\bh(i,j)|$ closest
        to $\lambda$. d) $\Delta b(i,j) = | \bh_g(i,j) - \bh_p(i,j) | / |\bh_g(i,j)|$ vs. the corresponding first $2000$
        coefficients of $| \Ah(i,j) |$ with the largest modulus. $ {\rm S/N} = 30~{\rm dB}$. }
        \label{fig:gcv_sat30_2}
\end{figure}

\begin{figure}
        \resizebox{\hsize}{!}{\includegraphics{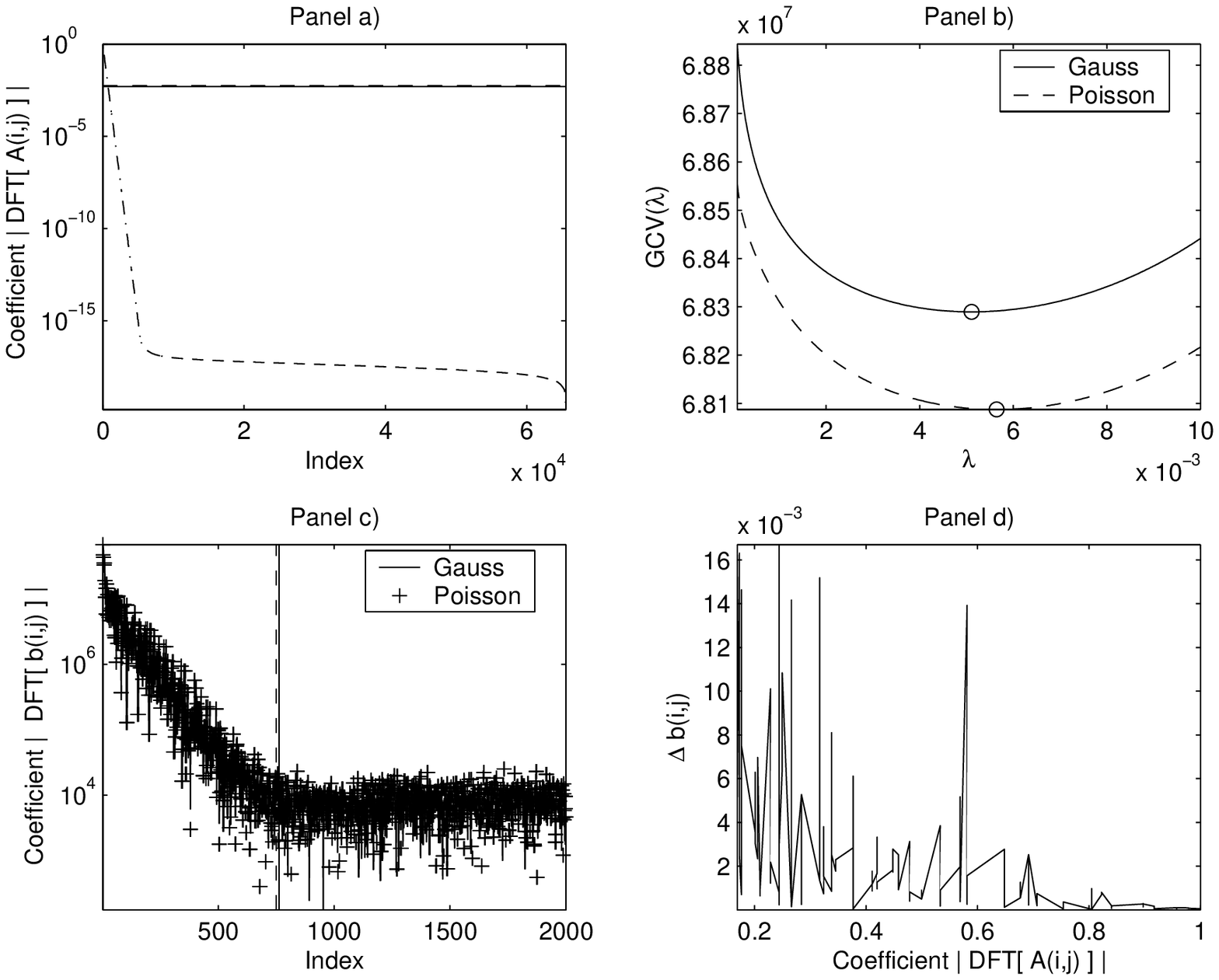}}
        \caption{Comparison of the results obtained by Tikhonov coupled with GCV in case of Poissonian
         and Gaussian noises, This figures correspond to the experiment shown in Fig.~\ref{fig:gcv_sat40_1}.
         Panel $a)$ the coefficients $| \bh_g(i,j) |$ and $| \bh_p(i,j) |$ in decreasing order.
        The two horizontal lines represent the values of $\lambda $ for the two noises; b) corresponding
        GCV functions; c) coefficients
        $|\bh_p(i,j)|$ and $|\bh_g(i,j)|$ corresponding to the first $2000$
        coefficients of $| \Ah(i,j) |$ shown in panel a). The vertical lines
        show the indices of $|\bh(i,j)|$ closest
        to $\lambda$. d) $\Delta b(i,j) = | \bh_g(i,j) - \bh_p(i,j) | / |\bh_g(i,j)|$ vs. the corresponding first $2000$
        coefficients of $| \Ah(i,j) |$ with the largest modulus. $ {\rm S/N} = 40~{\rm dB}$. }
        \label{fig:gcv_sat40_2}
\end{figure}

\begin{figure}
        \resizebox{\hsize}{!}{\includegraphics{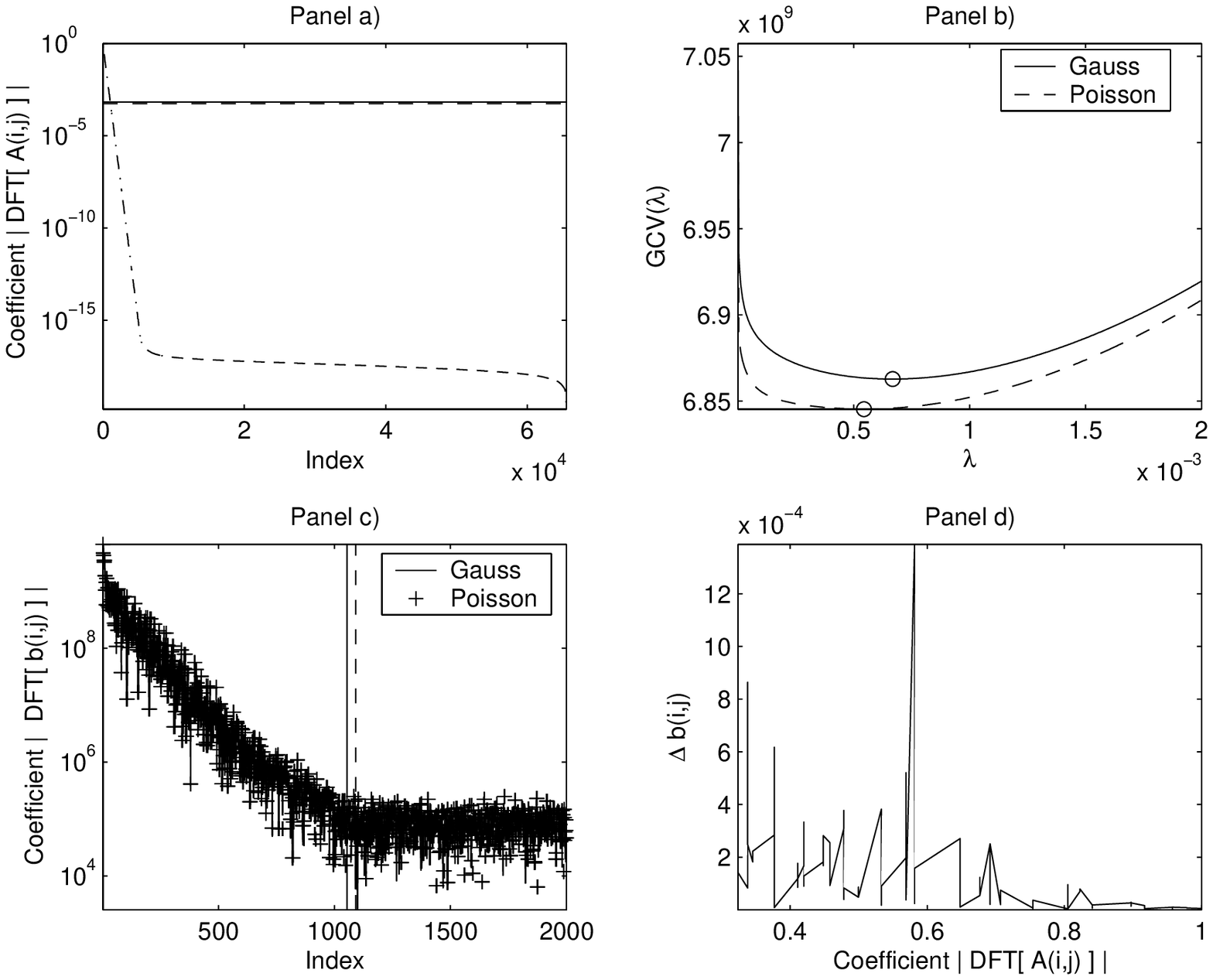}}
        \caption{Comparison of the results obtained by Tikhonov coupled with GCV in case of Poissonian
         and Gaussian noises, This figures correspond to the experiment shown in Fig.~\ref{fig:gcv_sat60_1}.
         Panel $a)$ the coefficients $| \bh_g(i,j) |$ and $| \bh_p(i,j) |$ in decreasing order.
        The two horizontal lines represent the values of $\lambda $ for the two noises; b) corresponding
        GCV functions; c) coefficients
        $|\bh_p(i,j)|$ and $|\bh_g(i,j)|$ corresponding to the first $2000$
        coefficients of $| \Ah(i,j) |$ shown in panel a). The vertical lines
        show the indices of $|\bh(i,j)|$ closest
        to $\lambda$. d) $\Delta b(i,j) = | \bh_g(i,j) - \bh_p(i,j) | / |\bh_g(i,j)|$ vs. the corresponding first $2000$
        coefficients of $| \Ah(i,j) |$ with the largest modulus. $ {\rm S/N} = 60~{\rm dB}$. }
        \label{fig:gcv_sat60_2}
\end{figure}

These results can be explained if one considers
Eq.~(\ref{eq:tikhonovf}), where it is clear that the role of
$\lambda$ is to replace the Fourier coefficients $\Ah(i,j)$ with
small modulus, i.e., those coefficients that make the deblurring
operation unstable. According to the two points mentioned above,
the ``optimal'' value of $\lambda$ should replace all the Fourier
coefficients whose modulus is smaller than the expected level of
the noise in the Fourier domain. Since in $b_p(i,j)$ and
$b_g(i,j)$ the level of the noise is the same, such a replacement
will be quite important and will have a similar effect for both
images. This is shown by the results in
Figs.~\ref{fig:gcv_sat20_2} - \ref{fig:gcv_sat60_2}. In
particular, the $c)$ panels show that GCV chooses $\lambda$ so
that the frequencies corresponding to the flat part of the spectra
(i.e., those dominated by the noise) are filtered out. The
consequence of this is that, for both Gaussian and the Poissonian
noises, almost the same number of coefficients are filtered.
Moreover, as is shown in the $d)$ panels, the coefficients
$|\bh_p(i,j)|$ and $|\bh_g(i,j)|$ corresponding to the
coefficients $\Ah(i,j)$ with the largest modulus, are very
similar. From this, one can conclude that the deblurred images
$\xb_{\lambda}$ can be expected to be very similar regardless of
the nature of the noise.

\begin{figure}
        \resizebox{\hsize}{!}{\includegraphics{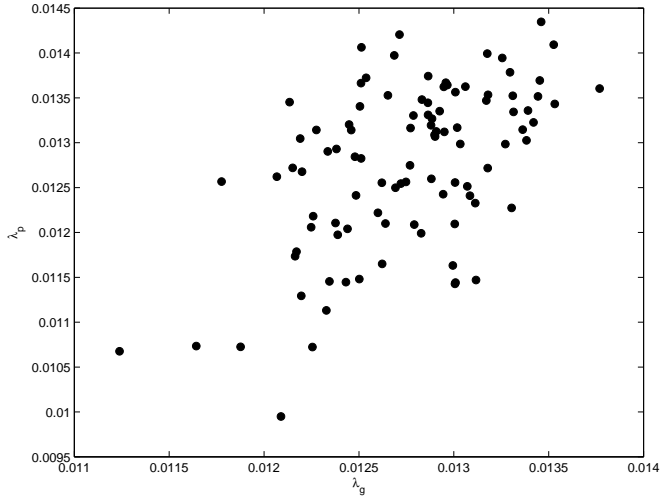}}
        \caption{Relationship between the estimates $\lambda_g$ and $\lambda_p$, corresponding to the
        Gaussian and the Poissonian noise, respectively, obtained in
        $100$ different realizations of the experiment shown in Fig.~\ref{fig:gcv_sat30_1}.
        The mean value and the standard deviation are $0.128$ and $4.55 \times 10^{-4}$ for $\lambda_g$,
        and $0.128$ and $9.10 \times 10^{-4}$ for $\lambda_p$.}
        \label{fig:fig_gcv_gp}
\end{figure}

The reason why the two GCV curves are almost identical (see the
$b)$ panels) is that, independently from the nature of the noise,
in Eq.~(\ref{eq:gcvf}) the quantity $\bh(i,j)$ can be replaced by
$\bh(i,j) = \Ah(i,j) \xh(i,j) + \zh_n(i,j)$, where $\zh_n(i,j)$ is
given by Eq.~(\ref{eq:noise}) or by a stationary white noise
process. Now, taking the expected value of the resulting ${\rm
GCV}(\lambda)$, it is not difficult to show that
\begin{align} \label{eq:gcva}
  {\rm E}[{\rm GCV}(\lambda)]  = &
                 N^2
                 \sum_{i,j=0}^{N-1}
                 \frac{|\Ah(i,j) \xh(i,j)|^2 + \sigma^2_{z_n}}
                       {(| \Ah(i,j) |^2+\lambda^{2})^2}
                  \nonumber \\
                  & /
                  \left|
                         \sum_{i,j=0}^{N-1}
                         \frac{1}
                              {|\Ah(i,j)|^2+\lambda^2}
                  \right|^2.
\end{align}
Since the term $\sigma_{z_n}^2$ is constant, the ${\rm E}[{\rm
GCV}(\lambda)]$ function is independent of the specific nature of
the noise. The same is not true for the variance. However, because
of the arguments presented above, no instabilities are to be
expected. This is supported by the fact that in our numerical
experiments we have never experienced stability problems (see also
Fig.~\ref{fig:fig_gcv_gp}).

\subsection{The iterative approach to regularization}
\label{sec:itappr}

Iterative algorithms are commonly used in deblurring problems.
Although, computationally less efficient than the direct methods,
such as the Tikhonov approach discussed above, they are much more
flexible in that they allow for the straightforward incorporation
of constraints. These algorithms provide regularization via a
semiconvergence property; that is, the iterates first reconstruct
the low frequency components of the signal, i.e. those less
contaminated by noise, and then the high frequency ones. In other
words, the number of iterations plays the same role as the
regularization parameter $\lambda$ in the Tikhonov approach.

Semiconvergence has been rigorously proved only for a limited
number of algorithms. For others, some theoretical results are
available but, the primary evidence stems from many years of
success in use on applied problems.

The prototypical iterative algorithm for least squares problems is
the {\it Landweber method} (LW). If $\xb_{0}$ is the starting image
(often $\xb_{0}=\zerob$), then the iterations take the form
\begin{equation} \label{eq:iter}
\xb_{k} = \xb_{k-1} + \omega \Ab^T \left[ \bb - \Ab
\xb_{k-1} \right],
\end{equation}
where, $k=1,2, \ldots $, and $\omega$ is a real positive parameter
satisfying $0 < \omega < 2 / \Vert \Ab^T \Ab \Vert$. The values of
$\omega$ determine, in part, the convergence of the iteration. The
semiconvergence property of this algorithm is typically proved
using arguments based on the singular values decomposition of the
matrix $\Ab$ \citep[for a discussion of this, see][]{vog02}.
However, it is, perhaps, more instructive to rewrite
Eq.~(\ref{eq:iter}) in the Fourier domain, obtaining
\begin{equation} \label{eq:lw_fourier}
\xh_{k}(i,j) = \frac{\bh(i,j)}{\Ah(i,j)}
\left[ 1 - (1 - \omega |\Ah(i,j)|^2)^k \right],
\end{equation}
with $0 < \omega < 2 / \max\left[|\Ah(i,j)|^2\right]$. If, as
usual, the PSF is assumed to have unit volume, then
$\max\left[|\Ah(i,j)|^2\right]=1$ and $0 < \omega < 2$. From this
equation, one can see that, for a given frequency index $(i,j)$,
the closer the term $\omega |\Ah(i,j)|^2$ is to one, the more
rapid is the convergence to $\bh(i,j) / \Ah(i,j)$, which
corresponds to the unregularized solution. Since, as mentioned
above, the largest values of the spectrum of $ \Ah(i,j) $
correspond to the lowest frequencies, it is evident from
\eqref{eq:lw_fourier} that the lower frequencies are restored in
early iterations, while progressively higher frequencies are
restored as the iteration progresses.

\subsubsection{Convergence properties}

\begin{figure}
        \resizebox{\hsize}{!}{\includegraphics{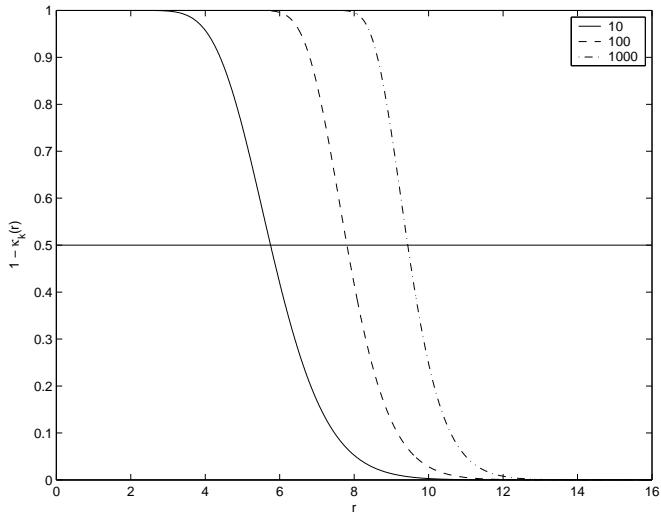}}
        \caption{Plot of $1-\kmath_k(r)$ vs. $r$ for different values of the iteration count $k$.}
        \label{fig:fig_iter_lw}
\end{figure}

Equation~(\ref{eq:iter}) shows that the convergence of LW
is driven by the rate with which the term
\begin{equation}
\kmath_k(i,j) = (1 - \omega | \Ah(i,j)|^2)^k
\end{equation}
goes to zero. In order to understand what this means in practical
situations, it is useful to see what happens in the case of a
noise-free image when the PSF is a two-dimensional Gaussian with
circular symmetry and dispersion $\sigma_p$. Without loss of
generality, we set $\omega = 1$. Then
\begin{equation}
\kmath_k(i,j) \approx \left[1 - \exp(- r^2 / \sigmah_p^2)\right]^k,
\end{equation}
where $r^2 = i^2 + j^2$ and $\sigmah_p$ is the dispersion of the
PSF in the frequency domain (see Eq.~(\ref{eq:sigmah})). From this
equation, it is not difficult to see that, even in case of
moderate values of $k$, the term within square brackets on the rhs
of Eq.~(\ref{eq:lw_fourier}) can be well-approximated by the
Boxcar function
\begin{equation}
\Pi_k(i,j) = \left\{ \begin{array}{ll}
1 & \quad \textrm{if $0 \leq | r | \leq r_{0.5,k}$}; \\
0 & \quad \textrm{otherwise}
\end{array} \right.,
\end{equation}
where $r_{0.5,k}$ is the value of $r$ for which $\kmath_k(r) =
0.5$ (see also Fig.~\ref{fig:fig_iter_lw}). Therefore, the
iterate~(\ref{eq:lw_fourier}) can be approximated by
\begin{equation} \label{eq:lw}
\xh_{k}(i,j) = \frac{\bh(i,j)}{\Ah(i,j)} \Pi_k(i,j).
\end{equation}

The requirement that $\kmath_k(i,j) \leq \epsilon$, with $0 < \epsilon < 1$
implies
\begin{equation} \label{eq:iterations}
k > \frac{\ln \epsilon}{\ln \left[1 - \exp(- r^2 / \sigmah_p^2)\right]}.
\end{equation}
This result shows that the restoration of the highest frequencies
(i.e., large $r$), requires a number of iterations that becomes
rapidly huge. For the case of the star-like sources, where all the
frequencies have to be restored, this means a terribly slow
convergence. More specifically, from Eq.~(\ref{eq:iterations}) one
can see that for frequencies $(i,j)$ such as $r \lessapprox
\sigmah_p$, $k \propto r$, while for larger values of $r$, $k$
increases exponentially. For example, some of the experiments
presented in this paper are based on images with size $256 \times
256$ pixels and with a Gaussian PSF with $\sigmah_p \approx 3.5$. 
In this case, if $\epsilon = 0.5$, in order to have
$r_{0.5,k}= \sigmah_p, 2 \sigmah_p, 3 \sigmah_p, 6 \sigmah_p$
(i.e., $r_{0.5,k} = 3.5, 7, 10.5, 21$), it is necessary that $k
\approx 2, 4, 5600, 3 \times 10^{15}$, respectively.

The obvious conclusion is that LW is useful only for the
restoration of objects for which the low-frequencies are dominant,
e.g. extended objects with smooth light distributions.

\subsubsection{Numerical results}

\begin{figure}
        \resizebox{\hsize}{!}{\includegraphics{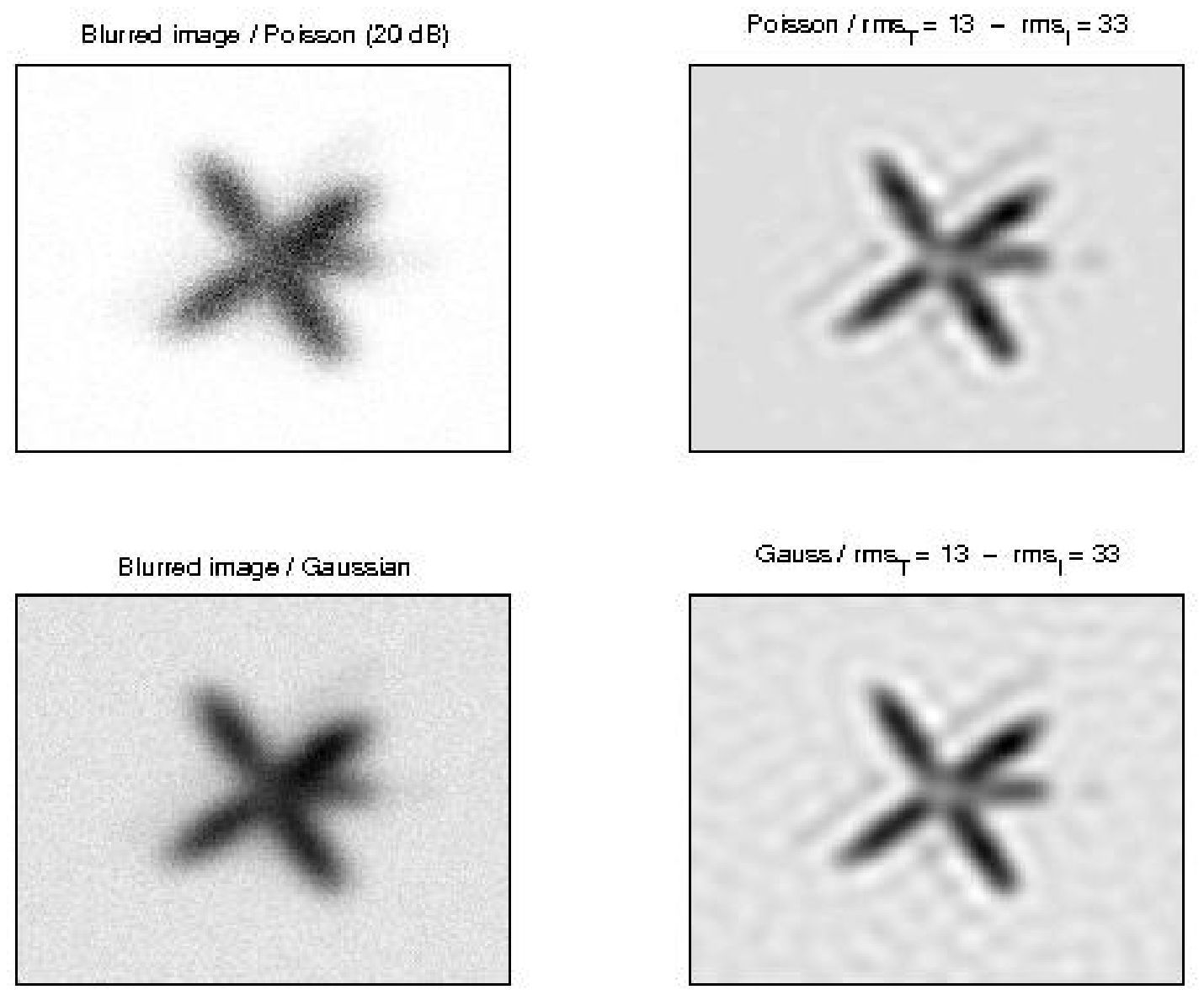}}
        \caption{LW deblurring of the images shown in Fig.~\ref{fig:gcv_sat20_1}.
         ${\rm rms}_T$ and ${\rm rms}_I$ are the rms of the true residuals calculated on the entire image and
         only on the pixels corresponding to the satellite, respectively.}
        \label{fig:gcv_sat20_3}
\end{figure}

\begin{figure}
        \resizebox{\hsize}{!}{\includegraphics{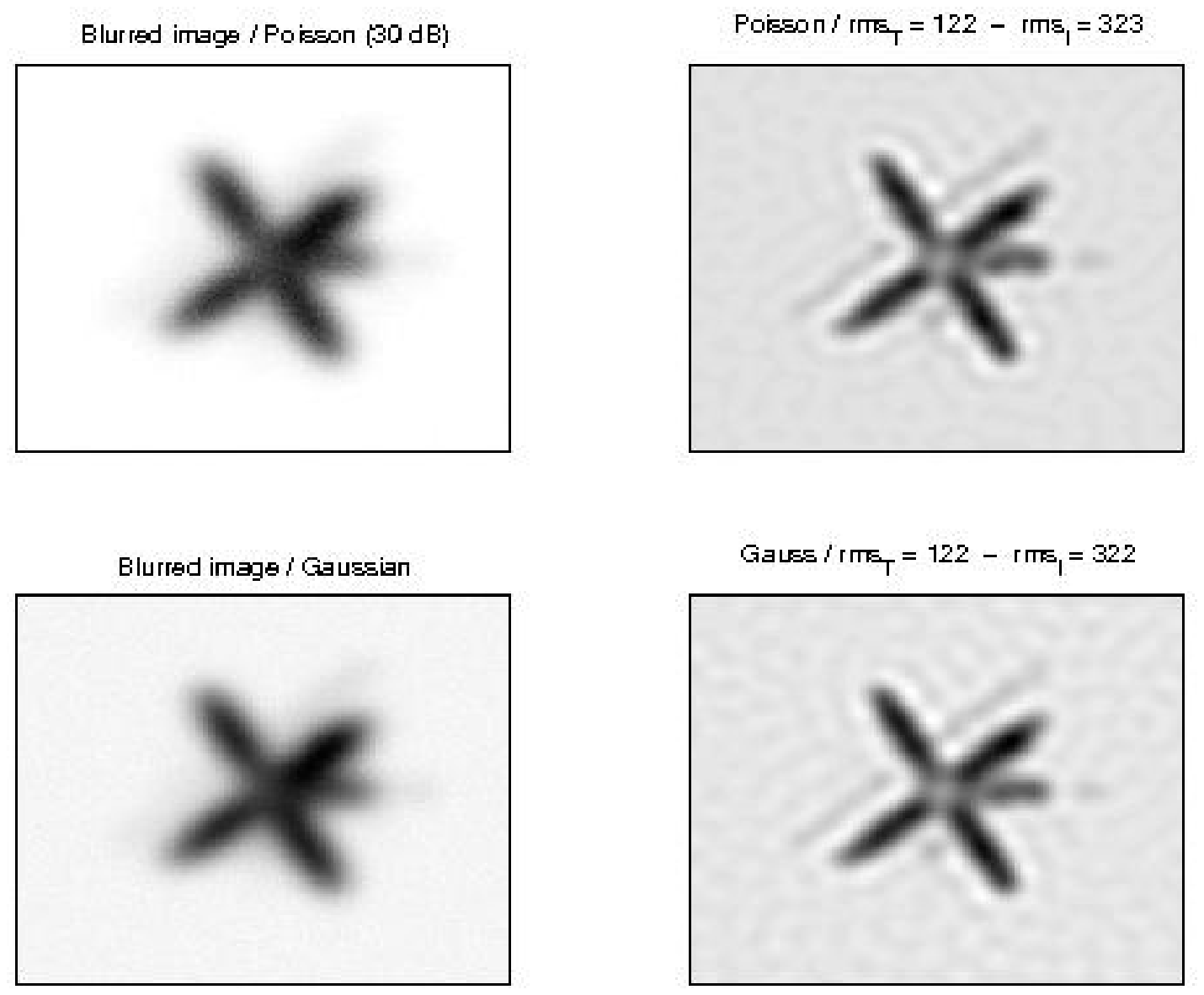}}
        \caption{LW deblurring of the images shown in Fig.~\ref{fig:gcv_sat30_1}.
         ${\rm rms}_T$ and ${\rm rms}_I$ are the rms of the true residuals calculated on the entire image and
         only on the pixels corresponding to the satellite, respectively.}
        \label{fig:gcv_sat30_3}
\end{figure}

\begin{figure}
        \resizebox{\hsize}{!}{\includegraphics{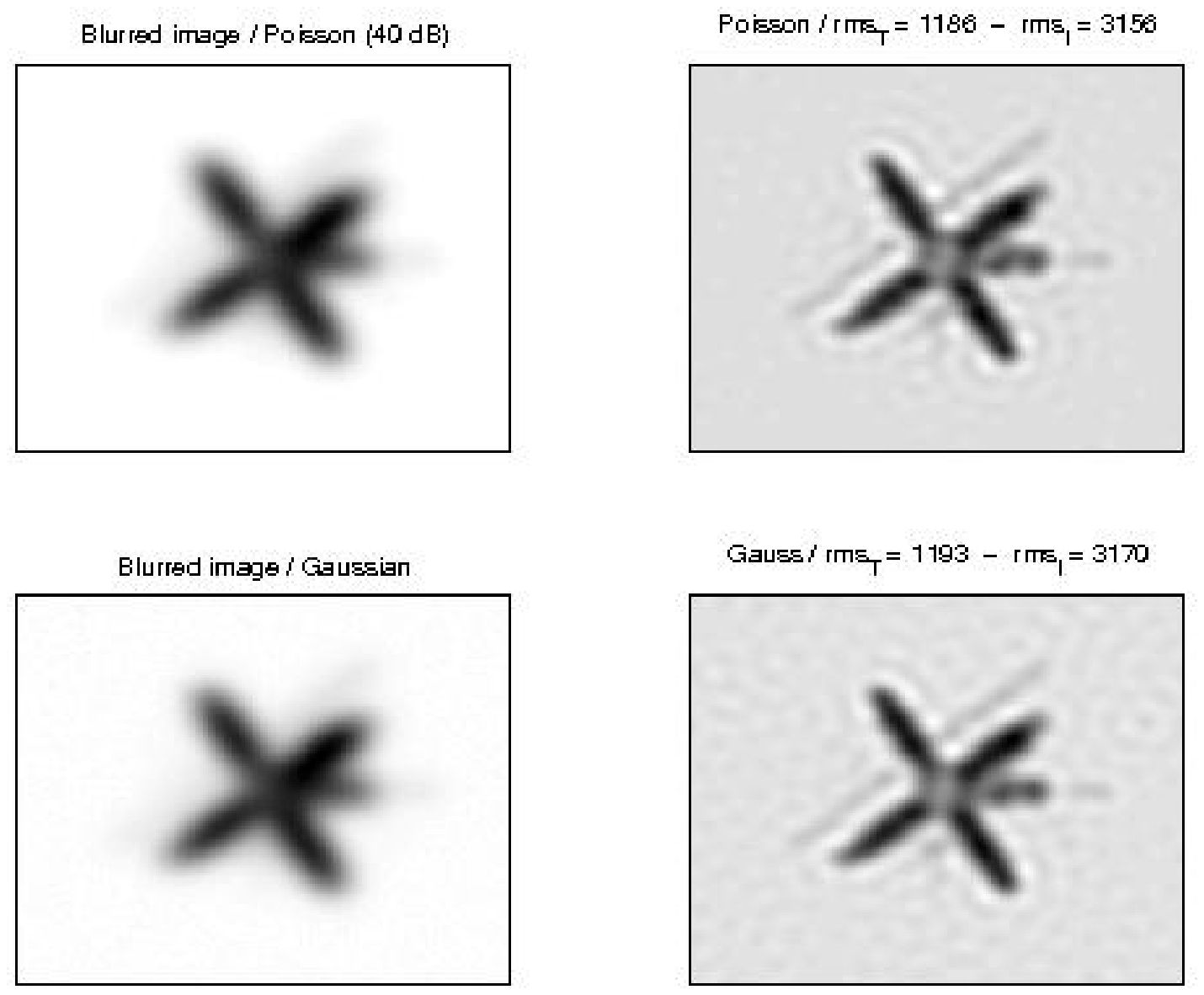}}
        \caption{LW deblurring of the images shown in Fig.~\ref{fig:gcv_sat40_1}.
         ${\rm rms}_T$ and ${\rm rms}_I$ are the rms of the true residuals calculated on the entire image and
         only on the pixels corresponding to the satellite, respectively.}
        \label{fig:gcv_sat40_3}
\end{figure}

\begin{figure}
        \resizebox{\hsize}{!}{\includegraphics{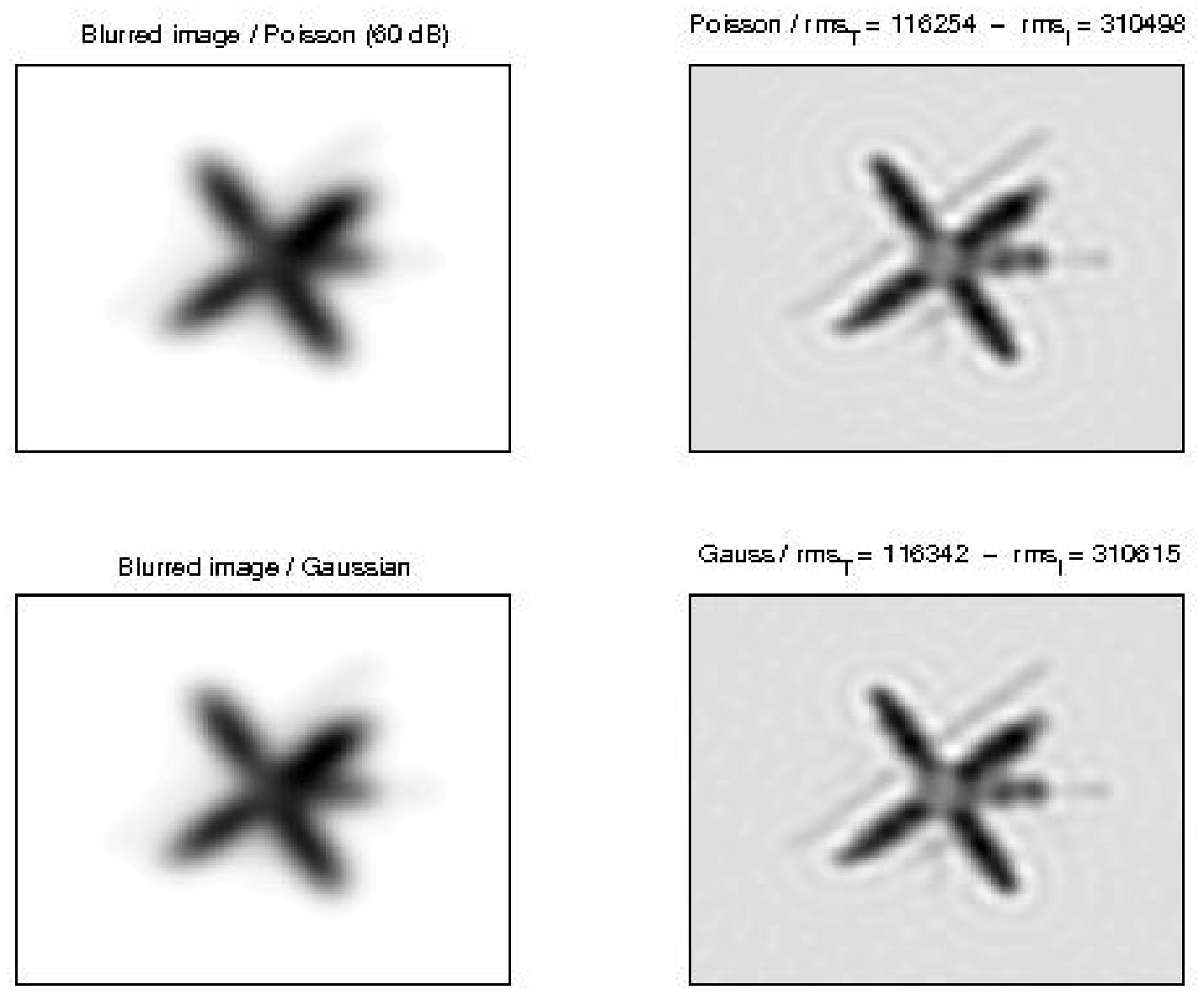}}
        \caption{LW deblurring of the images shown in Fig.~\ref{fig:gcv_sat60_1}.
         ${\rm rms}_T$ and ${\rm rms}_I$ are the rms of the true residuals calculated on the entire image and
         only on the pixels corresponding to the satellite, respectively.}
        \label{fig:gcv_sat60_3}
\end{figure}

\begin{figure}
        \resizebox{\hsize}{!}{\includegraphics{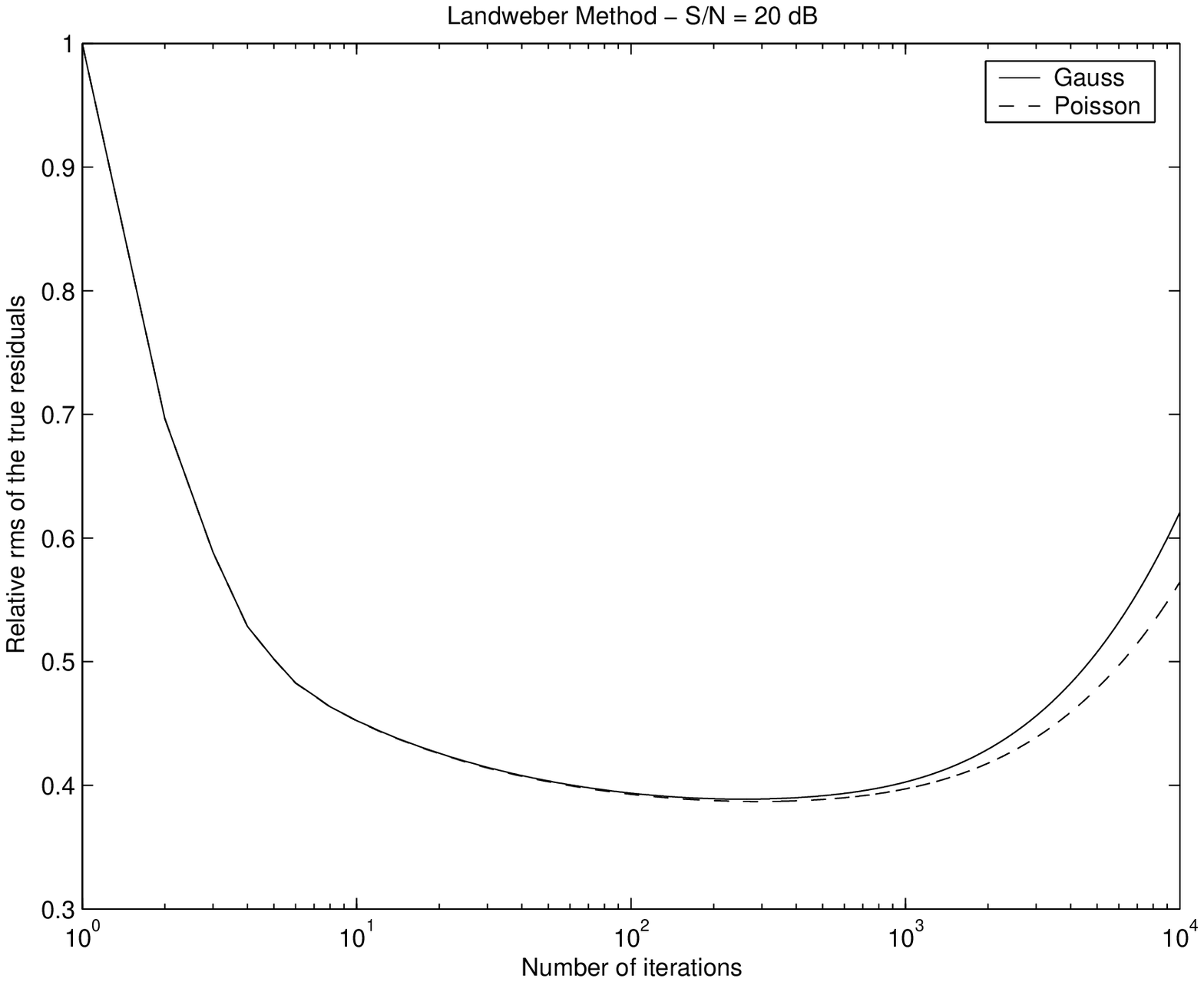}}
        \caption{Convergence rate of the LW algorithm applied to the problem of deblurring the images shown in
        Fig.~\ref{fig:gcv_sat20_1}.}
        \label{fig:gcv_sat20_4}
\end{figure}

\begin{figure}
        \resizebox{\hsize}{!}{\includegraphics{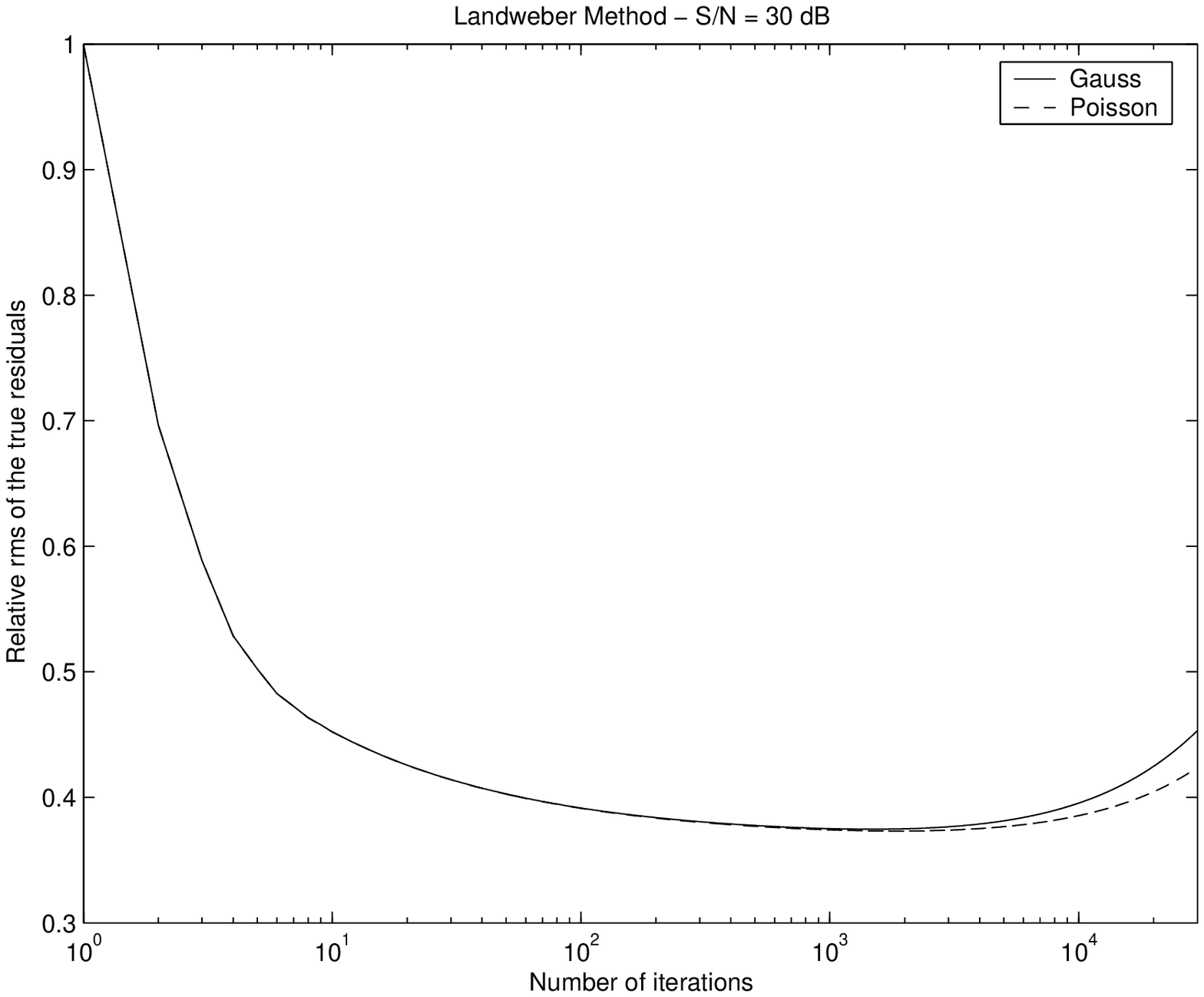}}
        \caption{Convergence rate of the LW algorithm applied to the problem of deblurring the images shown in
        Fig.~\ref{fig:gcv_sat30_1}.}
        \label{fig:gcv_sat30_4}
\end{figure}

\begin{figure}
        \resizebox{\hsize}{!}{\includegraphics{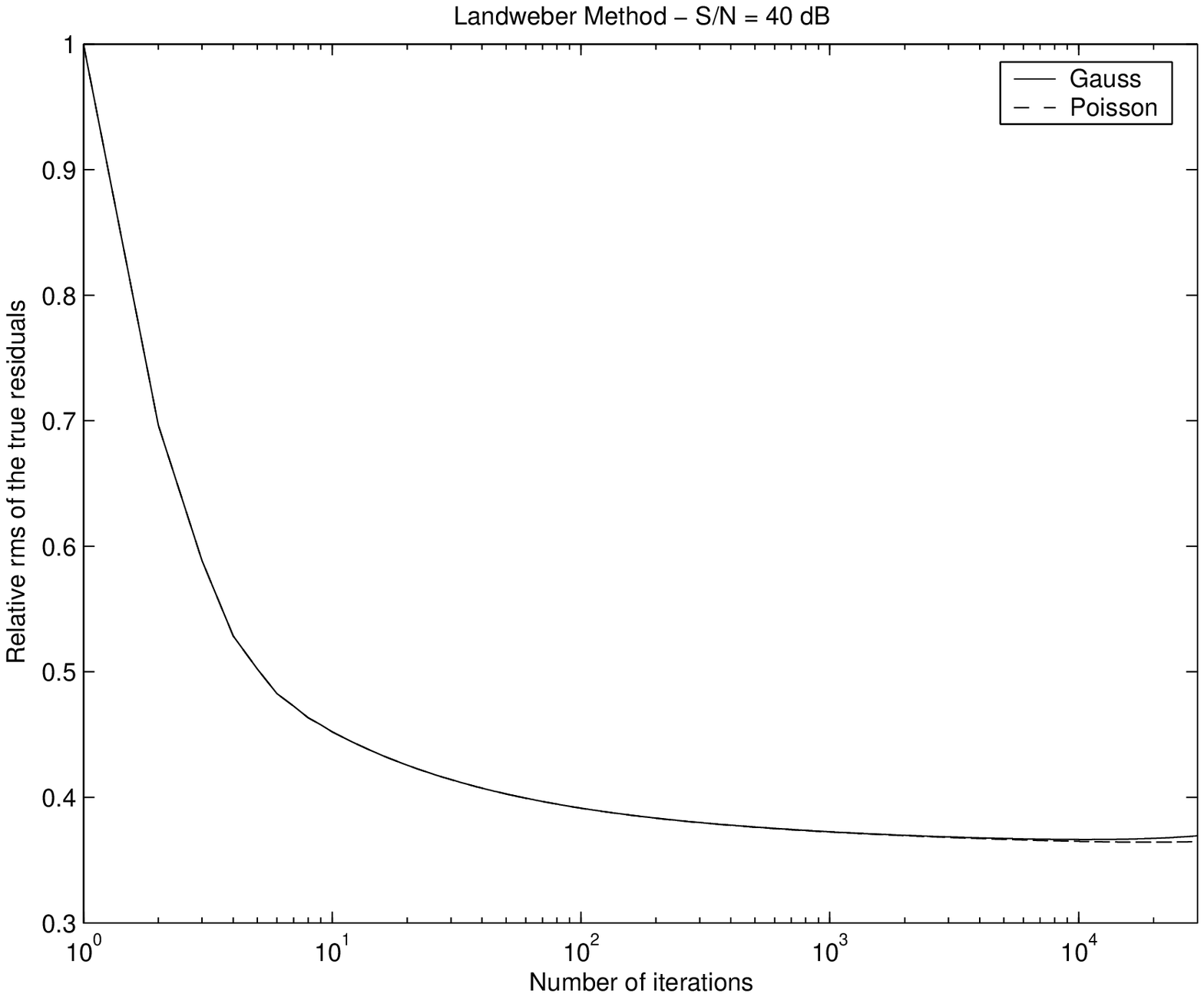}}
        \caption{Convergence rate of the LW algorithm applied to the problem of deblurring the images shown in
        Fig.~\ref{fig:gcv_sat40_1}.}
        \label{fig:gcv_sat40_4}
\end{figure}

\begin{figure}
        \resizebox{\hsize}{!}{\includegraphics{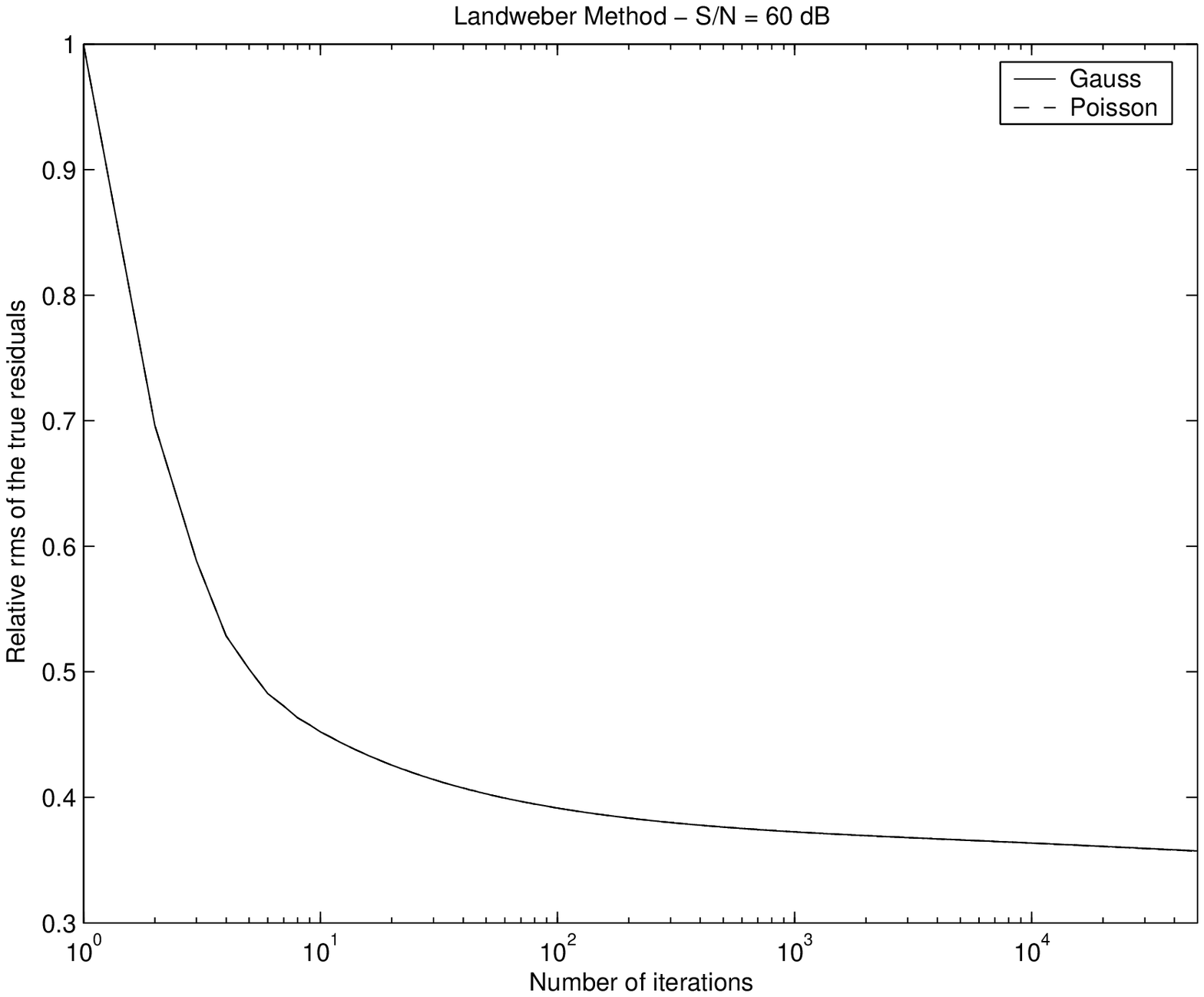}}
        \caption{Convergence rate of the LW algorithm applied to the problem of deblurring the images shown in
        Fig.~\ref{fig:gcv_sat60_1}.}
        \label{fig:gcv_sat60_4}
\end{figure}

Since, regardless of the noise type, LW reconstructs the lower
frequency components of the image first (i.e., the frequencies
where the contribution of the noise is negligible), we expect the
following for both $b_p(i,j)$ and $b_g(i,j)$:
\begin{enumerate}
\item the resulting deblurred images should be very similar; \\
\item in early iterations the convergence rate of the algorithms
should be almost identical.
\end{enumerate}
These statements are supported in
Figs.~\ref{fig:gcv_sat20_3}-\ref{fig:gcv_sat60_3} and
Figs.~\ref{fig:gcv_sat20_4}-\ref{fig:gcv_sat60_4}. In particular,
from the last set of figures one can see that the convergence
curves are almost identical until the minimum rms of the true
residual is reached. After that, because the high frequencies (the
ones that are more sensitive to the nature of the noise) begin to
be included in the restoration, the curves diverge.

\section{Richardson-Lucy Algorithm} \label{sec:rl}

In the previous sections we have shown that the LS methods are
relatively insensitive to the specific nature of the noise.
However, this does not mean that they are optimal. In principle,
methods that exploit the a priori knowledge of the statistical
characteristic of the noise should be able to provide superior
results.

In particular, model~(\ref{eq:stat3}) motivates the use of the
Richardson-Lucy (RL) algorithm for estimating $\xb$. RL can be
viewed as the EM algorithm corresponding to the statistical noise
model \eqref{eq:stat3}. The RL algorithm is defined by the
iteration
\begin{equation} \label{eq:iterrl}
\xb_{k+1} = \xb_k \odot \Ab^T\frac{\bb}{\bb_k},
\end{equation}
where $\bb_k = \Ab \xb_k$, and the fraction of two vectors denote
Hadamard (component-wise) division.

\begin{figure}
        \resizebox{\hsize}{!}{\includegraphics{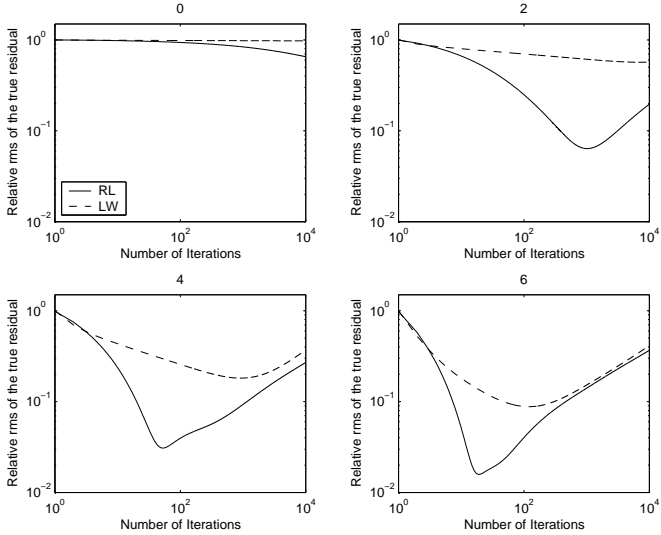}}
        \caption{$\Vert \xb_k - \xb \Vert / \Vert \xb \Vert$ vs. the number
        of iterations. The object of interest, located in the center of the image,
        is a two-dimensional Gaussian with circular
        symmetry and dispersion (in pixels) given in the top of each panel. It is superimposed to
        a background whose level is $1\%$ the peak value of the blurred image. The PSF is a
        two-dimensional Gaussian with a dispersion of $6$ pixels. The size of the image are $256 \times 256$
        pixels. The noise is Poissonian with
        peak ${\rm S/N} = 30~{\rm dB}$. Two deblurring algorithms are used: Richardson-Lucy (RL)
        and LW.}
        \label{fig:rllw_star_30_1}
\end{figure}

\begin{figure}
        \resizebox{\hsize}{!}{\includegraphics{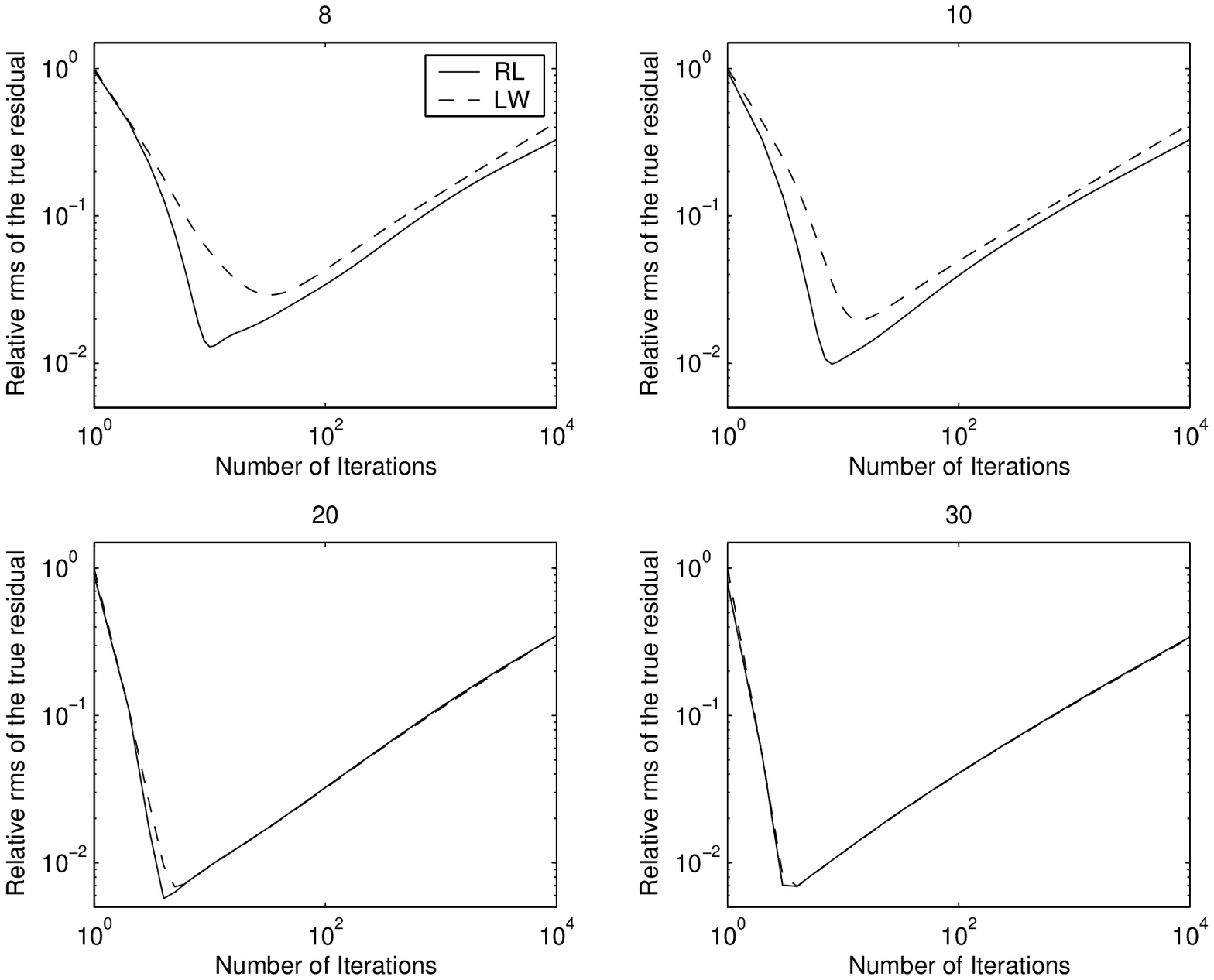}}
        \caption{As in Fig.~\ref{fig:rllw_star_30_1}.}
        \label{fig:rllw_star_30_2}
\end{figure}

\begin{figure}
        \resizebox{\hsize}{!}{\includegraphics{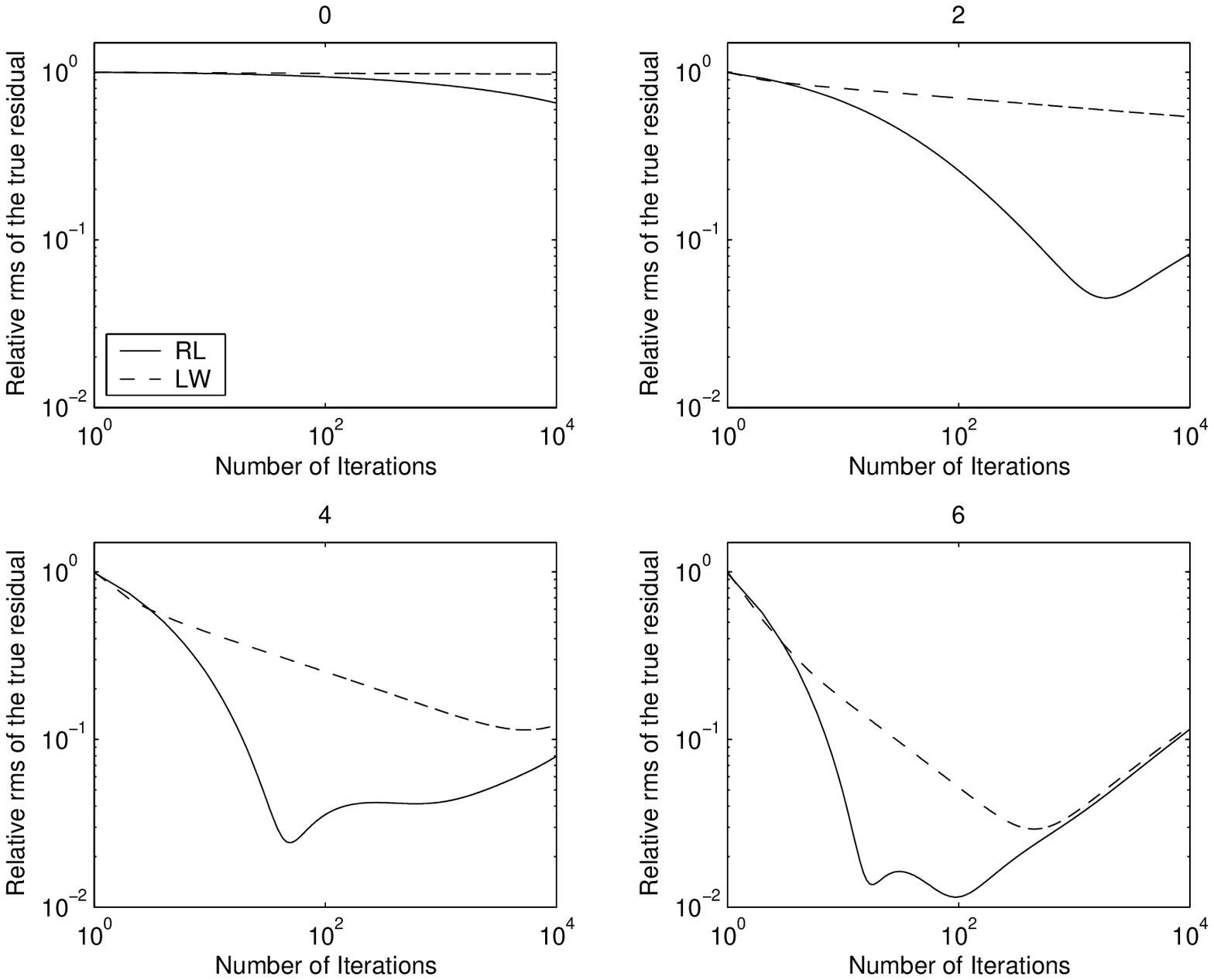}}
        \caption{As in Fig.~\ref{fig:rllw_star_30_1} but with ${\rm S/N} = 40~{\rm dB}$.}
        \label{fig:rllw_star_40_1}
\end{figure}

\begin{figure}
        \resizebox{\hsize}{!}{\includegraphics{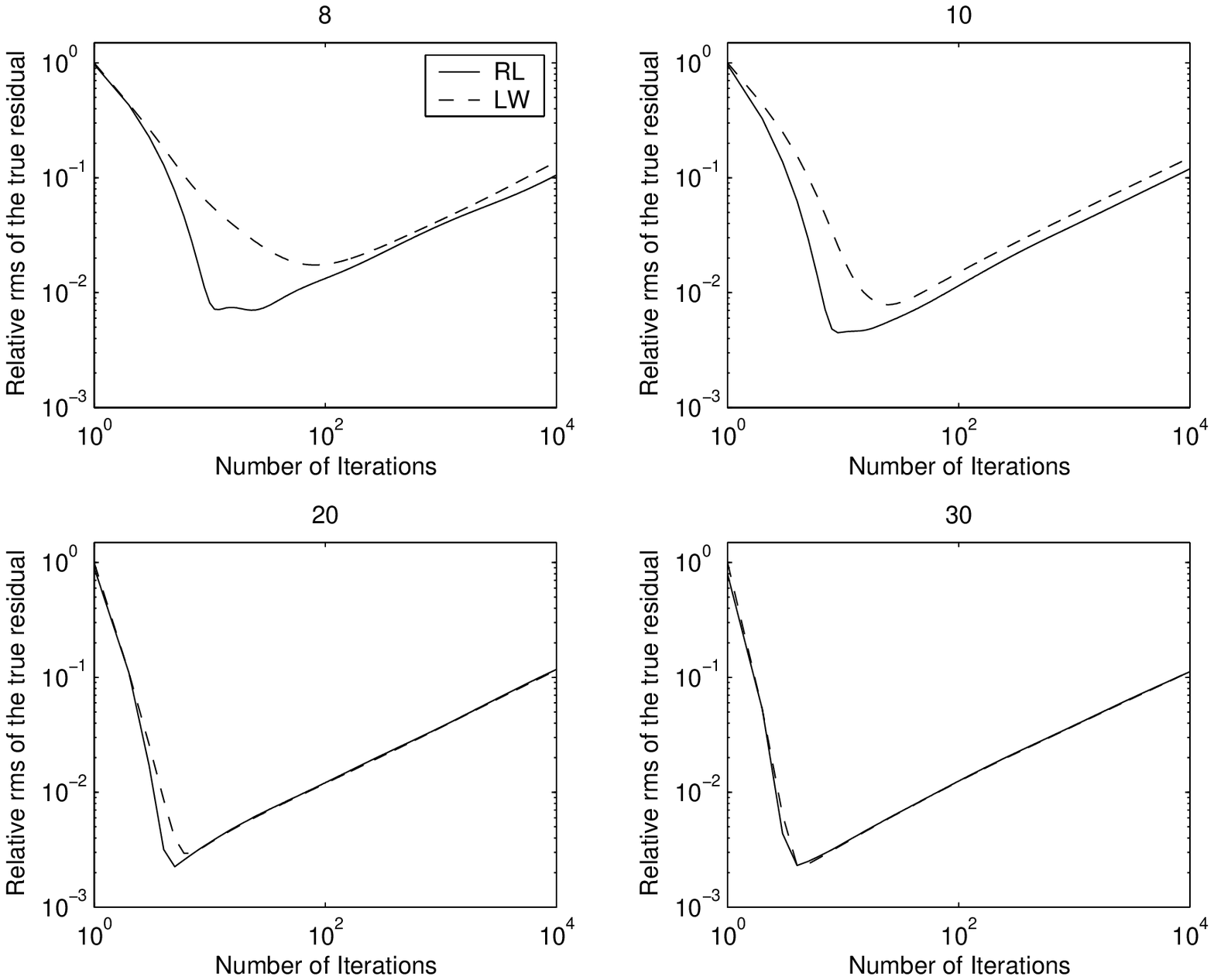}}
        \caption{As in Fig.~\ref{fig:rllw_star_30_1} but with ${\rm S/N} = 40~{\rm dB}$.}
        \label{fig:rllw_star_40_2}
\end{figure}

\begin{figure}
        \resizebox{\hsize}{!}{\includegraphics{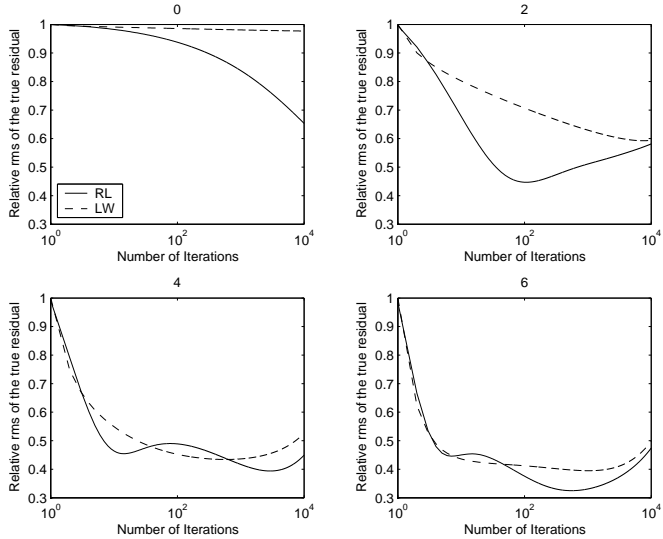}}
        \caption{$\Vert \xb_k - \xb \Vert / \Vert \xb \Vert$ vs. the number
        of iterations. The object of interest, located in the center of the image,
        is of a two-dimensional rectangular function,
        with side length (in pixels) given in the top of each panel. It is superimposed to
        a background whose level is $1\%$ the peak value of the blurred image. The PSF is a
        two-dimensional Gaussian with a dispersion of $6$ pixels. The size of the image are $128 \times 128$
        pixels. The noise is Poissonian with
        peak ${\rm S/N} = 30~{\rm dB}$. Two deblurring algorithms are used: Richardson-Lucy (RL)
        and LW.}
        \label{fig:rllw_rect_30_1}
\end{figure}

\begin{figure}
        \resizebox{\hsize}{!}{\includegraphics{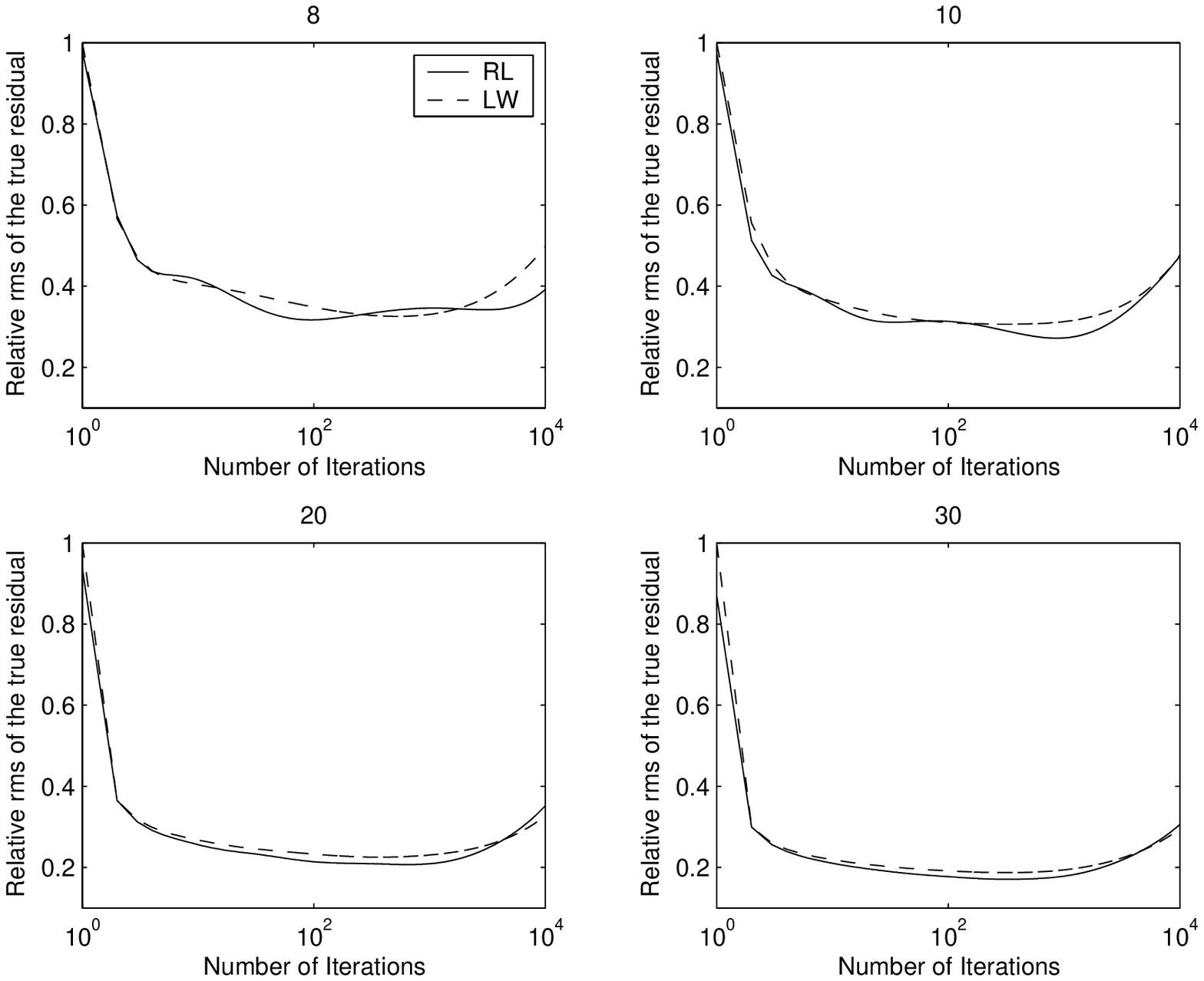}}
        \caption{As in Fig.~\ref{fig:rllw_rect_30_1}.}
        \label{fig:rllw_rect_30_2}
\end{figure}

\begin{figure}
        \resizebox{\hsize}{!}{\includegraphics{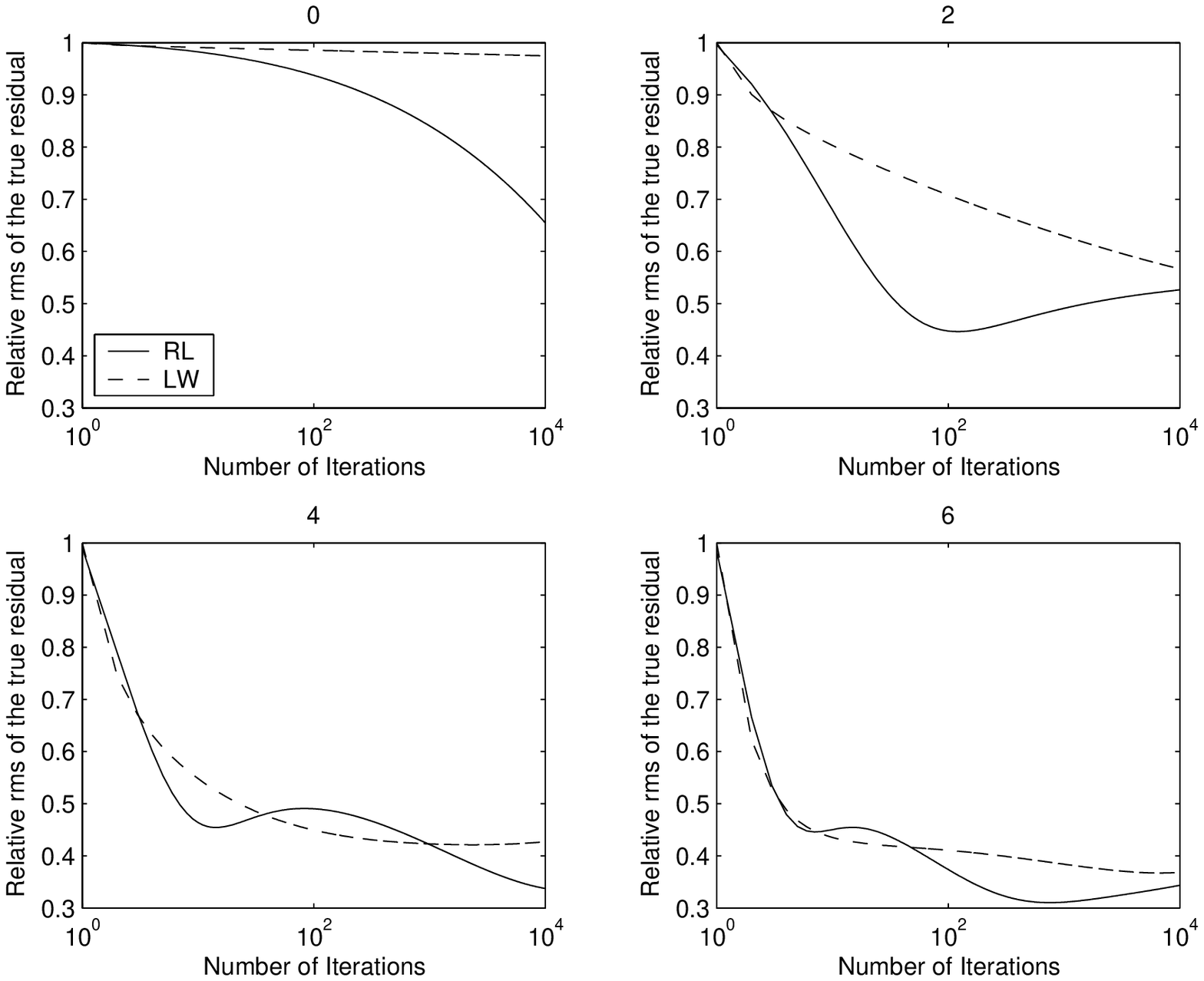}}
        \caption{As in Fig.~\ref{fig:rllw_rect_30_1} but with ${\rm S/N} = 40~{\rm dB}$.}
        \label{fig:rllw_rect_40_1}
\end{figure}

\begin{figure}
        \resizebox{\hsize}{!}{\includegraphics{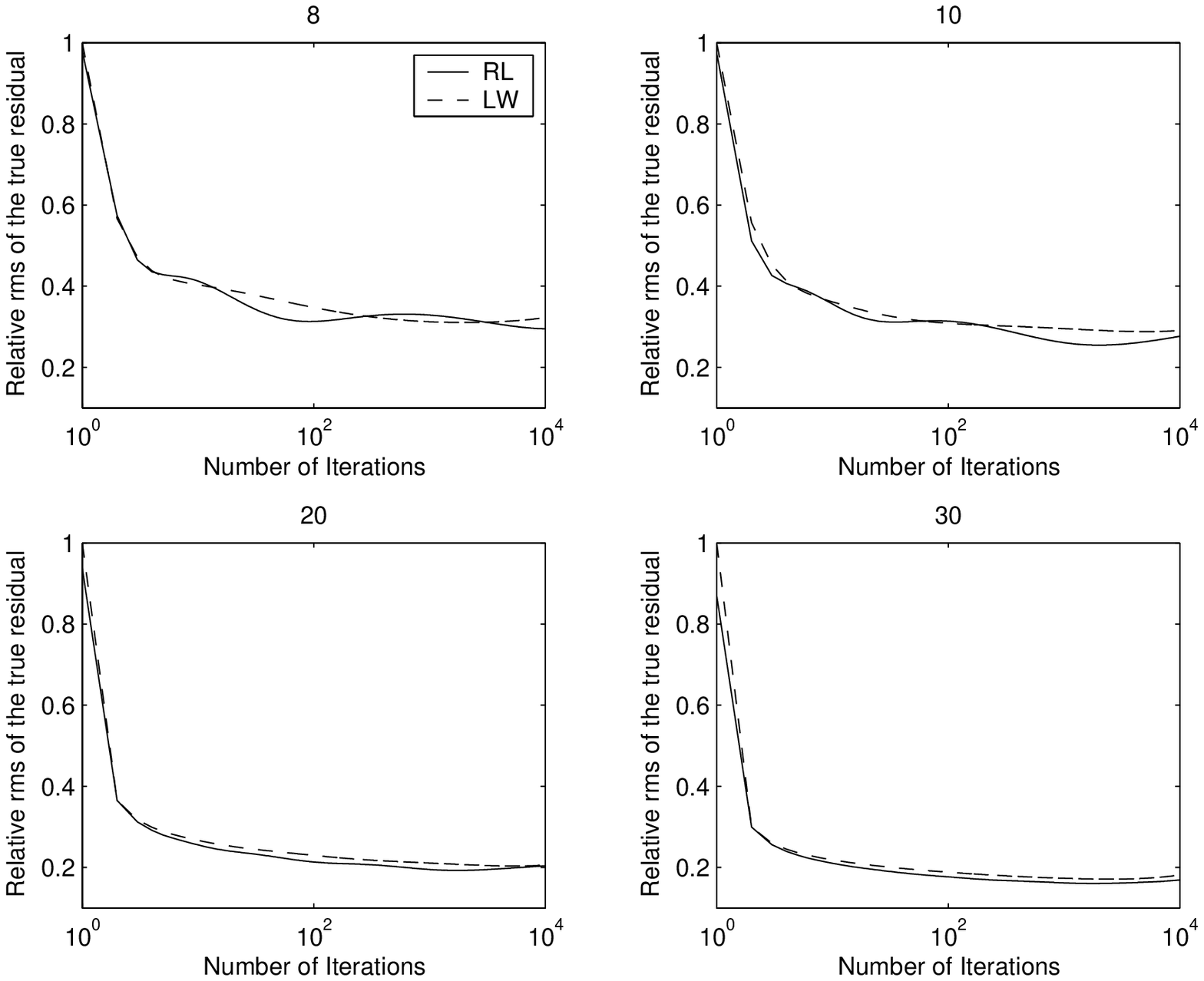}}
        \caption{As in Fig.~\ref{fig:rllw_rect_30_1} but with ${\rm S/N} = 40~{\rm dB}$.}
        \label{fig:rllw_rect_40_2}
\end{figure}

Since RL exploits the a priori knowledge regarding the statistics
of photon counts, it should be expected to yield more accurate
reconstructions than an approach that does not use this
information. In reality, as shown by
Figs.~\ref{fig:rllw_star_30_1}-\ref{fig:rllw_rect_40_2}, the
situation is not so clear. These figures provide the convergence
rates and the performances of RL and LW methods for objects with a
size that is increasing with respect to the size of the PSF (a
two-dimensional Gaussian with circular symmetry). Two different
types of objects are considered: a two-dimensional Gaussian and a
rectangular function. Since the first target object presents an
almost band-limited spectrum, whereas for the second target object
the high-frequency Fourier components are important, their
restorations represent very different problems. For both
experiments, a background with an intensity of $1\%$ of the
maximum value in the blurred image has been added. Finally, two
different levels of noise have been considered corresponding to a
peak ${\rm S/N}$ of $30$ and $40$ dB, respectively. The first case
provides a background with an expected number of counts
approximately equal to $30$, i.e., a level for which the Gaussian
approximation of the Poissonian distribution is not very good.

From Figs.~\ref{fig:rllw_star_30_1}-\ref{fig:rllw_rect_40_2} it
appears that the performance of RL for objects {\it narrower} than
the PSF is, in general superior to LW for the band-limited target.
The same is not true for the other objects. Interestingly, though,
for extended objects, i.e. smooth objects with high intensity
profiles over large regions, the performance of RL is roughly
equivalent to that of LS (to properly compare the convergence
rate, it is necessary to keep into account that, for each
iteration, RL requires the computation of twice the number of
two-dimensional DFT than is required by LW). This is especially
true for the images characterized by the best ${\rm S/N}$.
Motivated by these numerical results, we seek answers to the
following questions: (i) why does RL perform better than LS on
star-like objects, and (ii) why do the RL and LS approaches yield
roughly the same results on extended objects?

\subsection{RL vs LS: preliminary comments}
\label{sec:prelims}

It is important to note that in practice, when either the RL or LS
approaches are used in solving image reconstruction problems, the
exact computation of the maximum likelihood estimate (MLE) is not
sought. For example, as was
stated above, the LW iteration implements regularization
via the iteration count. In fact, the objective when using
LW is to stop the iteration late enough so that an accurate
reconstruction is obtained, but before the reconstruction is
corrupted by the noise in the high frequency components of the
image. Notice, for example, that in
Figs.~\ref{fig:rllw_star_30_1}-\ref{fig:rllw_rect_40_2} the
relative error begins to increase at a certain point in both the
RL and LW iterations.

As was stated above, one can show that the LW iterates are
regularized solutions of $\Ab^T\Ab\xb=\Ab^T\bb$ via the singular
value decomposition (SVD) of the matrix $\Ab$. Unfortunately, such
an analysis of RL is impossible due to the nonlinearity in the RL
algorithm. In particular, note the Hadamard multiplication and
division in algorithm \eqref{eq:iter}. Instead, we first note that
if $\Ab$ is an invertible matrix, RL iterates converge to the MLE
corresponding to the statistical model~ (\ref{eq:stat3})
\citep[see ][]{wu83}. Hence, RL can be viewed as an EM
algorithm \citep{car99}. The MLE is also the minimizer of the
negative log-likelihood function associated with
Eq.~(\ref{eq:stat3}); namely,
\begin{equation} \label{l_Poisson}
J(\xb) = \oneb^T\left[ \Ab\xb - \bb\odot\log(\Ab\xb)\right].
\end{equation}

When the RL algorithm is used on image deblurring problems it
exhibits a similar convergence behavior to that of the LW
iteration. Specifically, for ill-conditioned problems, the RL
iterates $\{\xb_k\}$ provide more accurate reconstructions in
early iterations (semiconvergence property), while in later
iterations blow-up occurs \citep{luc74, car99}. To
explain why this occurs, we note that the function $\Jb$ is
convex. In fact, assuming $A$ is positive definite, $\Jb$ is
strictly convex. In this case, the minimum of $\Jb$ is the unique
solution of the equation $\nablab \Jb(\xb)=\zerob$, where $\nablab
\Jb$ is the gradient $\Jb$; that is,
\begin{equation} \label{eq:grad_eqn}
\Ab^T \left(\oneb - \frac{\bb}{\Ab\xb}\right)=\zerob.
\end{equation}
It is clear that provided $\Ab$ is invertible, the solution of
\eqref{eq:grad_eqn} is obtained when $\Ab\xb=\bb$. That is,
$\xb_*=\Ab^{-1}\bb$ is the solution of Eq.~(\ref{eq:grad_eqn}).
Under the same conditions on $\Ab$, the LW iteration
converges to the same value. Thus, we would expect that since
blow-up occurs as $\xb_k\rightarrow \xb_*$ in the LW
iteration when $\Ab$ is a poorly conditioned matrix, we will see
the same results when we use RL. One consequence of this fact is
that during reconstruction, RL uses, effectively, only a few
frequencies and therefore cannot fully exploit prior statistical
information regarding the noise (this should require the use of
the entire spectrum).

\subsection{RL vs. LS: a sensitivity analysis}

In order to obtain a deeper understanding of the RL and LS
approaches and to answer the two questions posed above,
it is useful to introduce the
quantities
\begin{align}
\Deltab_{{\rm LW}}(\xb_k) & = - \Ab^T \rb_k; \label{eq:dellw} \\
\Deltab_{{\rm RL}}(\xb_k) & = - \xb_k \odot \Ab^T \frac{\rb_k}{\Ab
\xb_k}, \label{eq:delrl}
\end{align}
which provide the correction to the solution $\xb_{k}$ at the k-th
iteration for LW and RL, respectively. Here,
\begin{equation}
\rb_k = \Ab \xb_k - \bb,
\end{equation}
and, without loss of generality, we have set $\omega=1$ in the LW
iteration. We note that in order to obtain Eq.~(\ref{eq:delrl}) we
needed the identity $\Ab^T \oneb = \oneb$ to hold. From these
equations it is evident that at each iteration LW corrects the
solution $\xb_k$ with a quantity proportional to $\rb_k$, while
the correction provided by RL is proportional to $\xb_k$ itself.
Thus it is not surprising that RL outperforms LW when applied to
reconstructing objects composed of star-like sources on a flat
back ground, since in the early stages of both iterations the
entries of $\xb_k$ are large and increasing in regions
corresponding to the positions of the objects in the image, while
the values of $\rb_k$ are correspondingly small and decreasing.

However, as has been shown by the simulations presented above, RL
does not outperform LW when applied to reconstructing objects
with smooth light distribution and
whose spatial extension is broader than the PSF. In order to
understand this phenomena, it is useful consider the negative
Jacobian matrices of the quantities~(\ref{eq:dellw}) and
(\ref{eq:delrl}):
\begin{equation}
\label{eq:JLW} J_{LW}(\xb_k)= \Ab^T \Ab
\end{equation}
and
\begin{align}
\label{eq:JRL} J_{RL}(\xb_k) &= \hbox{diag}\left[ \Ab^T
\frac{\rb_k}{\Ab\xb_k}\right] \nonumber\\
& + \hbox{diag}[\xb_k] ~
\Ab^T\hbox{diag}\left[\frac{\bb}{\Ab\xb_k\odot
\Ab\xb_k}\right]\Ab.
\end{align}
These matrices provide the sensitivities of the LW and RL
algorithms to changes in the components of the iterate $\xb_k$.
Equations \eqref{eq:JLW} and \eqref{eq:JRL} allow us to make
several observations. We begin by considering \eqref{eq:JLW}.
Since in general astronomical applications $\Ab$ is the
discretization of a PSF with an almost limited spatial support,
the LW sensitivity matrix $\Ab^T\Ab$ will also have spatial
support that is almost limited. From this observation, we can
conclude that for a given pixel the corresponding component of the
vector $\Deltab_{\rm LW}$ will be most sensitive to changes in the
value of the pixel itself and in the values of ``nearby" pixels.
Here, the term ``nearby" is defined by the characteristics of the
PSF. More specifically, as the spread of the PSF increases, so
does the collection of ``nearby" pixels.

Perhaps an even more important observation, is that the
sensitivity of the LW iteration to perturbations in $\xb_k$
is independent of both $\xb_k$ and $\bb$. Consequently, the
algorithm has no means of distinguishing between low and high
intensity regions within the object, and hence, perturbations of
the same magnitude are allowed for components of $\xb_k$
corresponding to regions of both low and high light intensity.
This explains why in areas of low light intensity (where the
margin of error is very small) LW, and the least squares
approach in general, does poorly.

The sensitivity matrix \eqref{eq:JRL} for the RL iteration is more
difficult to analyze. However, some simplification is possible
when one considers the problem of the restoration of a flat
background or of regions of an image in which the intensity
distribution varies smoothly (e.g., the interior of the
rectangular function considered in the simulations). In fact, in
this case it is possible to define a region $\Omega$ where the
image can be considered constant or almost constant. Because of
the semiconvergence property of RL, in such regions the components
of the vector $\rb_k$ converge rapidly to zero (this has been
verified via numerical simulations). Thus the first term in
\eqref{eq:JRL} converges to zero. The same is not true for the
second term. Thus, it is reasonable to expect that it will provide
an accurate approximation of the sensitivity of RL within
$\Omega$.

Provided that the spread of the PSF is small
relative to the size of $\Omega$, early in RL iterations the vector $\xb_k$ is
approximately constant and close to $\bb$, i.e. those pixels
values are reconstructed rapidly, within $\Omega$. Hence, the
vector $\bb/(\Ab\xb_k\odot\Ab\xb_k)\approx 1/\Ab\xb_k$ is also
approximately constant within $\Omega$. In addition, if we define
$\Db_\Omega$ to be the diagonal matrix with components
\begin{equation}
  \label{D_i}
  [\Db_\Omega]_{jj} = \left\{ \begin{array}{ll}
    1, & j\in \Omega \\
    0, & j\notin \Omega
  \end{array} \right.,
\end{equation}
then
\begin{equation}
\label{eq:approx} \Db_\Omega \Ab\xb_k\approx \Db_\Omega \Ab
\Db_\Omega \xb_k
\end{equation}
will be accurate within the interior of $\Omega$. To obtain
\eqref{eq:approx} we used the fact that $\xb_k$ is approximately
constant on $\Omega$ and that the spread of $\Ab$ is small
compared to the size of $\Omega$. Finally, the second term of
\eqref{eq:JRL} can be approximated within $\Omega$ as follows:
\begin{align}
\Db_\Omega & \hbox{diag}[\xb_k]    \Ab^T \hbox{diag}[\bb/(\Ab\xb_k\odot \Ab\xb_k)]
                        \Ab \nonumber \\
                        &\approx \hbox{diag}[\xb_k]  \Db_\Omega \Ab^T
                        \Db_\Omega\hbox{diag}[\bb/(\Ab\xb_k\odot \Ab\xb_k)]
                                              \Ab \label{eq:approx1}\\
                        &\approx \hbox{diag}[\Db_\Omega(\xb_k/\Ab\xb_k)] \Ab^T
                        \Ab\label{eq:approx2}\\
                        &\approx \Db_\Omega \Ab^T
                        \Ab.\label{eq:approx3}
\end{align}
Approximation \eqref{eq:approx1} follows from \eqref{eq:approx}.
Approximation \eqref{eq:approx2} follows from the fact that, as
stated above, early in RL iterations
$\bb/(\Ab\xb_k\odot\Ab\xb_k)\approx 1/\Ab\xb_k$ is approximately
constant. Thus we see that not only does the second term in
\eqref{eq:JRL} not converge to zero, it is well-approximated
within $\Omega$ by the LW sensitivity \eqref{eq:JLW}.
Recalling that the first term in \eqref{eq:JRL} converges rapidly
to zero in RL iterations, it is therefore not surprising that RL
and LW provide similar results in the interior of the
rectangular object mentioned above. We can extend this discussion
to extended objects in general by noting that such objects can be
decomposed into a union of regions in which the light intensity is
approximately constant. Hence, RL and LW should provide
similar results for extended objects in general.

\subsection{RL vs. LS: convergence properties}

As shown above, LW presents an acceptable convergence rate only in
case of restoration of extended objects. Unfortunately,
understanding the convergence properties of the RL algorithm
\eqref{eq:iter} is a very difficult affair since it is not
possible to carry out an analysis similar to that done in
Sect.~\ref{sec:itappr}. For this reason, in order to obtain some
insight, we consider, again, a noise-free signal $\bb$ and
Gaussian PSF with circular symmetry and variance $\sigma^2_p$. In
addition, we suppose that the object of interest is a circular
Gaussian source with variance $\sigma^2_b$. The amplitude of the
source is not considered since RL conserves the total number of
counts. Due to the connection between the RL and LW iterations
discussed in the previous section, an understanding of RL
convergence may provide further understanding of the convergence
of the LW iteration. For simplicity, in what follows we work in
the continuous, and results will be later discretized.

If a Gaussian function with variance
$\sigma^2$ is denoted by $G[\sigma^2]$, the following facts are useful:
\begin{align}
\frac{G[\sigma^2_1]}{G[\sigma^2_2]} & =  G\left[\frac{\sigma^2_1
\sigma^2_2}{\sigma^2_2 - \sigma^2_1}\right];
\label{eq:oper1} \\
G[\sigma^2_1] \odot G[\sigma^2_2] & =  G\left[\frac{\sigma^2_1
\sigma^2_2}{\sigma^2_2 + \sigma^2_1}\right];
\label{eq:oper2} \\
G[\sigma^2_1] \otimes G[\sigma^2_2] & =  G\left[ \sigma^2_1 +
\sigma^2_2 \right]. \label{eq:oper3}
\end{align}
Here, the symbol ``$~\otimes$'' indicates convolution. From these
equations it is evident that the result of any of the above
operations produces a new Gaussian function. Only the first
operation requires a condition be satisfied, i.e., $\sigma^2_2 >
\sigma^2_1$. This will always be satisfied during the RL iteration
(see Eq.~(\ref{eq:ineq1}) below).

A point worth noting is that if we define $\sigma^2$ to be the
variance of the Gaussians on the right hand side of
\eqref{eq:oper1}-\eqref{eq:oper3}, then for equation
\eqref{eq:oper1} we have $\sigma^2_1 < \sigma^2$; in equation
\eqref{eq:oper2} we have $\sigma^2 < \sigma^2_1 < \sigma^2_2$; and
in equation \eqref{eq:oper3} we have $\sigma^2 = \sigma^2_1 +
\sigma^2_2$. Consequently, only the operation~(\ref{eq:oper2})
results in a Gaussian function with a variance that is smaller
than both $\sigma_1$ and $\sigma_2$.

If the true object $\xb$ is a Gaussian with variance
$\sigma^2_o$, then using \eqref{eq:oper3} and the fact that
$\Ab\xb=\bb$, it is
\begin{equation}
\label{sigma_o} \sigma^2_b =\sigma^2_o + \sigma^2_p.
\end{equation}
As an initial guess in the RL algorithm, we take $\xb_0 = \oneb$.

Now, let's suppose that $\xb_k = G[\sigma^2_k]$. Using Eqs.~
(\ref{eq:oper1}) - (\ref{eq:oper3}) one can obtain
\begin{equation} \label{eq:sigma_k}
G[\sigma^2_{k+1}] = G\left[\frac{\sigma^2_k (\sigma^2_k \sigma^2_b
+ \sigma^4_p + \sigma^2_k \sigma^2_p)} {(\sigma^2_k +
\sigma^2_p)^2} \right].
\end{equation}
It is a straightforward exercise to show that
\begin{equation} \label{eq:ineq1}
R(\sigma^2_k) = \frac{\sigma^2_{k+1}}{\sigma^2_k} = \frac{\sigma^2_k
\sigma^2_b + \sigma^4_p + \sigma^2_k \sigma^2_p}
{(\sigma^2_k + \sigma^2_p)^2} < 1
\end{equation}
provided
\begin{equation} \label{eq:ineq2}
\sigma^2_k > \sigma^2_b - \sigma^2_p.
\end{equation}
To prove that \eqref{eq:ineq2} holds for all $k$, we use
induction. First, note that since $\xb_0=\oneb$,
$\Ab^T\oneb=\oneb$, and $\Ab^T=\Ab$, we have that
$\xb_1=\Ab^T\bb=G[\sigma_p^2 + \sigma_b^2]$. Then
$\sigma_1^2=\sigma_p^2 + \sigma_b^2$, and hence,
Eq.~(\ref{eq:ineq2}) is satisfied for $k=1$. Now, we show that if
Eq.~(\ref{eq:ineq2}) holds for $k$, it must hold also for $k+1$.
By replacing $\sigma_{k+1}^2$ in Eq.~(\ref{eq:ineq2}) by the
argument of the Gaussian function on the right hand side of
Eq.~(\ref{eq:sigma_k}), one can obtain an equivalent inequality
involving $\sigma_k^2$ given by
\begin{equation} \label{eq:ineq3}
q(\sigma_k^2)>0,
\end{equation}
where $q(\sigma^2)$ defined by
\begin{eqnarray*}
q(\sigma^2) &=& \sigma^2 (\sigma^2 \sigma^2_b +
          \sigma^4_p+\sigma^2 \sigma^2_p) + (\sigma_p^2-\sigma_b^2)
          (\sigma^2 + \sigma^2_p)^2\\
     &=& 2\sigma_p^2(\sigma^2)^2 + (3\sigma_p^4 -
         2\sigma_p^2\sigma_b^2)\sigma^2 +
         (\sigma_p^6-\sigma_b^2\sigma_p^4).
\end{eqnarray*}
Notice that $q$ is a quadratic function. We can therefore find its
zeros via the quadratic formula. These are given by
\begin{equation}
\sigma^2 = -\frac{\sigma_p^2}{2},\;\sigma_b^2-\sigma_p^2.
\end{equation}
Since $\sigma_p^2>0$, we know that the graph of $q$ is an upward
opening parabola. Furthermore, by \eqref{sigma_o} we have
$\sigma_b^2-\sigma_p^2=\sigma_o^2>0$, and hence, we know that if
$\sigma^2>\sigma_b^2-\sigma_p^2$, then $q(\sigma^2)>0$. Thus
\eqref{eq:ineq3} follows from the inductive hypothesis, and our
proof is complete.

In light of these findings, it is possible to consider some convergence
properties of the RL algorithm. We begin by showing that the
sequence $\{\sigma_k^2\}$ converges to
$\sigma_o^2=\sigma_b^2-\sigma_p^2$. First, note that
Eqs.~(\ref{eq:ineq1}) and (\ref{eq:ineq2}) imply that
$\{\sigma_k^2\}$ is a decreasing sequence that is bounded below by
$\sigma_b^2-\sigma_p^2$. Hence, $\{\sigma_k^2\}$ converges to some
$\sigma^*\geq \sigma_b^2-\sigma_p^2$. From inequalities
\eqref{eq:ineq1} and \eqref{eq:ineq2}, we have that
$R(\sigma_*^2)=1$. Furthermore, the arguments used in the proof of
\eqref{eq:ineq2} imply that if $\sigma^2>\sigma_b^2-\sigma_p^2$
then $R(\sigma^2)<1$. Thus it must be that
$\sigma_*^2=\sigma_b^2-\sigma_p^2$.

\begin{figure}
        \resizebox{\hsize}{!}{\includegraphics{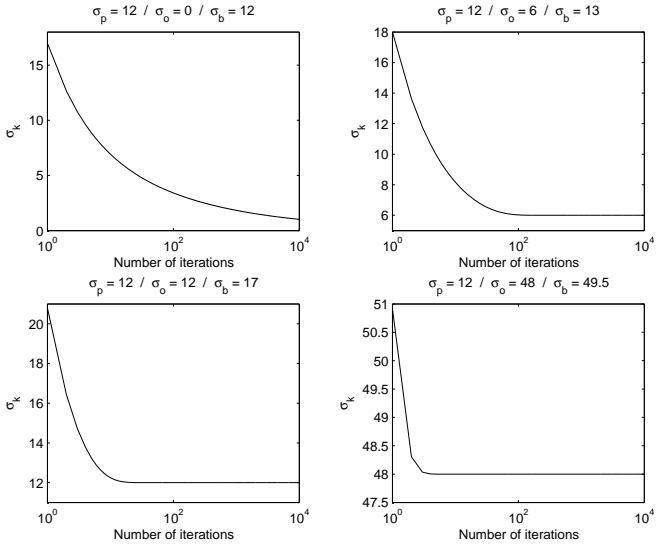}}
        \caption{$\sigma_k$ vs. the number of iterations for RL in the case of a Gaussian source with
        various values of $\sigma_o$ and a Gaussian PSF with dispersion $\sigma_p=6$.}
        \label{fig:fig_iter_1}
\end{figure}

\begin{figure}
        \resizebox{\hsize}{!}{\includegraphics{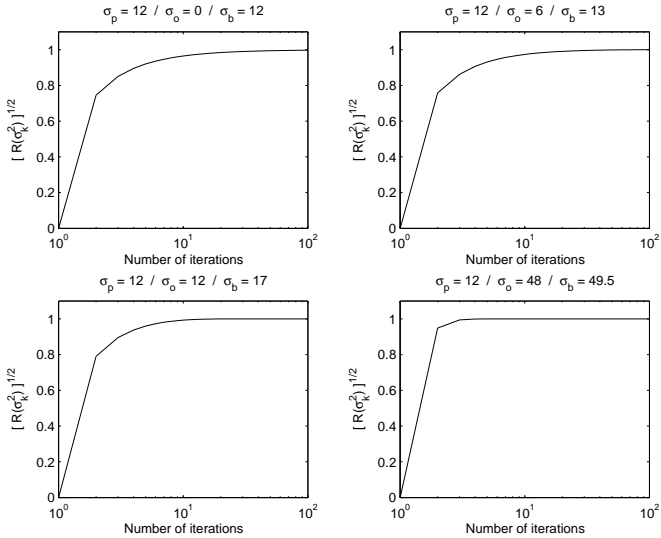}}
        \caption{$R^{1/2}(\sigma_k^2)$ vs. the number of iterations for the cases shown in Fig.~\ref{fig:fig_iter_1}.}
        \label{fig:fig_iter_2}
\end{figure}

In regard to the convergence rate of the RL algorithm,
Eq.~(\ref{eq:ineq1}) shows that, almost independently from the
characterists of the object, in the very first iteration, when
$\sigma_b \gg \sigma_p$, we have
\begin{equation}
R(\sigma^2_0) = \frac{\sigma_0^2 (\sigma_b^2 + \sigma_p^2)}{\sigma_0^4} \approx 0.
\end{equation}
In fact, if $\xb_{\zerob}=\oneb$ (i.e., $\sigma_0^2 = \infty$),
it is not difficult to see that $\xb_1 = \Ab \bb$, i.e., the result
of the first iteration is given by $G(\sigma_p^2+\sigma_b^2)$.
At this point, there are two possible situations:
\begin{enumerate}
\item  For extended objects, we have $\sigma^2_b \approx
\sigma^2_1\gg \sigma^2_p$. In this case, $R(\sigma^2_1) \approx
1$. In general, this means that we can expect rapid progress in
early iterations; after that the convergence rate slows down
remarkably. This behavior is similar to that of the LW algorithm;

\item For stellar-like objects, we have $\sigma^2_b \approx
\sigma^2_p$. Now, if we set $\sigma_k = \alpha ~\sigma_p$, then
\begin{equation}
R(\sigma^2_k) = \frac{1 + 2 \alpha^2}{(1+\alpha^2)^2}.
\end{equation}
For example, when $\alpha=1,0.5, 1/3, 1/6$, then $R(\sigma^2_k)=0.750, 0.960, 0.990, 0.999$,
respectively.
In other words, although the convergence rate of RL slows
down as the iteration progresses, this effect is not as pronounced
as it is for the LW algorithm.
\end{enumerate}
These statements are confirmed by Figs.~\ref{fig:fig_iter_1},
\ref{fig:fig_iter_2}.

A comparison between the RL solution for star-like sources at the
k-th iterate
\begin{equation} \label{eq:comp1}
\xh_k(i,j) \propto \exp{(- r^2 / 2 \sigmah_k^2)}
\end{equation}
with the corresponding LW solution
\begin{equation} \label{eq:comp2}
\xh_k(i,j) = \Pi_k(i,j),
\end{equation}
provides some additional insight into the convergence properties
of these algorithms. In fact, from Eq.~(\ref{eq:comp1}) it is
evident that although the high frequencies are filtered in the RL
algorithm, the filter is less stringent for high frequencies than
is the Landweber filter. The consequence is that, in general, 
at a given $k$, RL has available a broader range of frequencies 
to restore the object. On the one hand, this can improve the
convergence rate of RL compared to LW; on the other hand this
could create problems when one or more star-like objects are superimposed with
an extended object. In fact, a few RL iterations are sufficient to
restore the extended object. The same is not true for the
star-like objects. Therefore, more iterations are necessary. However,
because of the amplification of the noise, this means the degradation
of the results in the parts of the image not occupied by the star-like 
objects.

\begin{figure}
        \resizebox{\hsize}{!}{\includegraphics{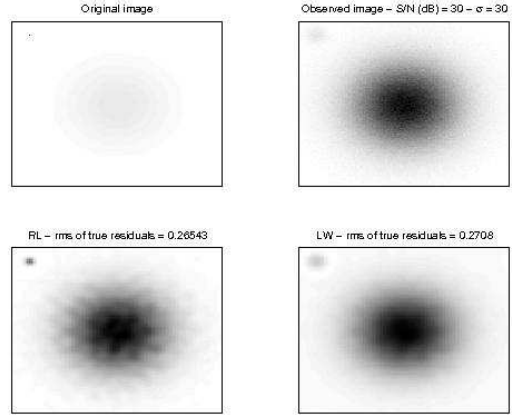}}
        \caption{Deblurring results obtained with RL and LW for a star-like object
        superimposed with an extended source given by a Gaussian with circular symmetry and dispersion set to
        $30$ pixels. An additional background is present with a level set to $1\%$ the maximum of the noise-free
        blurred image. The PSF is a Gaussian with circular symmetry and dispersion set to $6$ pixels.
        The size of the image are $128 \times 128$ pixels. The noise is Poissonian with peak
        ${\rm S/N} = 30~{\rm dB}$. The image presented corresponds to the result with the smallest rms of the
        true residuals.}
        \label{fig:fig_comparison}
\end{figure}

 This effect is clearly visible in the
experiment shown in Fig.~\ref{fig:fig_comparison}.

\section{Conclusions} \label{sec:conclusions}

In this paper we provide explanations for why, in spite the
incorporation of a priori information regarding the noise
statistics of image formation, the RL deblurring algorithm often
does not provide results that are superior to those obtained by
techniques based on an LS approach. In particular, we have
identified a possible explanation in the regularization approaches
of the specific algorithms. In fact, the adoption of a priori
smoothness constraint in the Tikhonov approach, or the need to
stop the iterations before blow-up occurs in the iterative
approaches, e.g. both LW and RL, do not permit the full
exploitation of the information contained in the highest Fourier
frequencies, i.e., those where the specific nature of the noise
has the largest influence. This has two consequences: I) the
performance of the LS algorithms is almost insensitive to whether
noise is Gaussian or Poissonian; II) the RL algorithm does not
fully benefit from the fact that it incorporates the specific
statistical model of the noise. In other words, the regularization
of the solution implies a levelling out of the possible
performances. In this respect, much more than a detailed knowledge
of the nature of the noise is needed. Specifically, some rough a
priori information regarding the solution, e.g. is it an extended
or star-like object, is needed before one can know whether or not
RL will provide superior results. Our numerical experiments
support these conclusions. In particular, the fact that
reconstructions gotten via the RL algorithm are often comparable
to those of LW, i.e., an unsophisticated and very slow algorithm,
indicates that resorting to advanced and often complex techniques
is not always justified.

We stress that such conclusions are not only of academic interest.
In fact, with respect to the ML algorithms, in general the LS
algorithms are much easier to implement, are more flexible as
concerns the incorporation of constraints, are more amenable to a
theoretical analysis of their characteristics and are
computationally less costly. Consequently, unless the use of a
different approach is justified, they should be considered the
standard approach.

\clearpage
\clearpage

\clearpage

\clearpage

\clearpage

\clearpage

\clearpage

\clearpage

\clearpage

\end{document}